\documentclass[aps,prd,eqsecnum]{revtex4}

\usepackage{graphicx}
\usepackage{lscape}
\usepackage{indentfirst}
\usepackage{latexsym}
\usepackage{multirow}
\usepackage{tabls}
\usepackage{epsfig}
\usepackage{color}
\usepackage{amssymb}

\usepackage{amsfonts}
\usepackage{amsmath}
\usepackage{bm}
\usepackage{mathrsfs}

\def\arreq{ \!\! = \!\! }

\def\barh{ {\bar h} }

\def\bj{ {\bf j} }

\def\bk{ {\bf k} }
\def\bn{ {\bf n} }

\def\bv{ {\bf v} }

\def\bx{ {\bf x} }

\def\hath{ {\hat{h}} }

\def\bzero{ {\bm 0} }

\def\lsim{\mathrel{\rlap{\lower3pt\hbox{\hskip1pt$\sim$}}
    \raise1pt\hbox{$<$}}}                
\def\gsim{\mathrel{\rlap{\lower3pt\hbox{\hskip1pt$\sim$}}
    \raise1pt\hbox{$>$}}}         

\def\coordeq{ \, \mathrel{ \rlap{\hbox{\hskip-2.5pt$=$} }
    \raise4pt\hbox{$\cdot$}} \, }                

\begin{document}

\title{Radiation reaction and gravitational waves in the effective field theory approach}

\author{Chad R. Galley\footnote{crgalley@umd.edu} and Manuel Tiglio\footnote{tiglio@umd.edu}}


\affiliation{Center for Scientific Computation and Mathematical Modeling and\\
Center for Fundamental Physics, Department of Physics, University of Maryland, College Park, MD, 20742}


\begin{abstract}
We compute the contribution to the Lagrangian from the leading order (2.5 post-Newtonian) radiation reaction and the quadrupolar gravitational waves emitted from a binary system using the effective field theory (EFT) approach of Goldberger and Rothstein. We use an initial value formulation of the underlying (quantum) framework to implement retarded boundary conditions and describe these real-time dissipative processes. We also demonstrate why the usual scattering formalism of quantum field theory inadequately accounts for these. The methods discussed here should be useful for deriving real-time quantities (including radiation reaction forces and gravitational wave emission) and hereditary terms in the post-Newtonian approximation (including memory, tail and other causal, history-dependent integrals) within the EFT approach.  We also provide a consistent formulation of the radiation sector in the equivalent effective field theory approach of Kol and Smolkin.
\end{abstract}

\maketitle

\section{Introduction} \label{sec:intro}

The anticipated detection of gravitational waves using ground-based laser interferometers \cite{Abramovici:LIGO, Giazotto:VIRGO, Willke:GEO, Ando:TAMA} has spurred significant advances towards a detailed theoretical 
description of binary compact objects. From a semi-analytical point of view, a major tool for understanding these systems 
comes from the post-Newtonian (PN) expansion, which is a perturbation theory based on the small relative velocity $v$ of the binary's constituents. Chief among these accomplishments are the heroic calculations of the equations of motion for the orbital dynamics through 3.5PN order and the power emitted in gravitational waves, which are estimated to be sufficiently accurate for precisely estimating the parameters of the detected gravitational wave sources. See \cite{Blanchet:LRR, FutamaseItoh:LRR} for reviews of the standard treatments of the PN approximation and their main results.

The recently introduced effective field theory (EFT) approach of Goldberger and Rothstein \cite{GoldbergerRothstein:PRD73}, called non-relativistic general relativity (NRGR), provides an economical way to apply the PN approximation to compact binaries.
The NRGR approach is viable in any system wherein the wavelengths of gravitational perturbations are much larger than the sources or scatterers. It is a valuable tool that utilizes the efficient diagrammatic and regularization techniques of perturbative quantum field theory to calculate quantities of interest within the post-Newtonian expansion. Only tree-level Feynman diagrams are relevant for (classical) astrophysical systems so that quantum corrections are completely ignorable. NRGR was further developed in  \cite{Porto:PRD77} and \cite{GoldbergerRothstein:PRD73_2} to incorporate spin effects and dissipation from
gravitational wave absorption, respectively. It has also been recently used to derive the 2PN equations of motion for
nonspinning compact objects in \cite{GilmoreRoss:0810.1328} using Kol-Smolkin variables \cite{KolSmolkin:CQG25} to streamline the number of contributing Feynman diagrams. NRGR has also been used in \cite{Leibovich:NRGRtalk} to derive the radiation reaction force on an electrically charged extended body interacting with its own electromagnetic radiation (see Section \ref{sec:other} for discussion pertaining to the applicability and consistency of the particular method used in \cite{Leibovich:NRGRtalk}). See \cite{Goldberger:LesHouches} for a good review of NRGR and also \cite{CannellaSturani:0808.4034, SanctuarySturani:0809.3156, ChuGoldbergerRothstein:JHEP3, Chu:0812.0012} for applications beyond the two-body problem. In addition, NRGR has provided the first derivation of the 3PN spin-spin potential \cite{PortoRothstein:PRL97, Porto:grqc0701106, PortoRothstein:PRD78, PortoRothstein:PRD78_2} and of a quadrupole-spin correction to the gravitational energy \cite{Porto:PRD73}.

NRGR uses the language and formalism of perturbative quantum field theory via path integral methods. However, only the classical limit of the theory is relevant for describing astrophysical compact binary systems so that all quantum effects are blatantly ignored. Nevertheless, there are different path integral formalisms that can be used in NRGR to implement different boundary conditions on the gravitational perturbations. The boundary conditions are typically chosen to conform to the questions being asked. The ``in-out" formalism, which is the framework typically discussed in many quantum field theory texts and was used in \cite{GoldbergerRothstein:PRD73} to introduce NRGR, is based on a scattering formalism that uses the Feynman Green's function to implement scattering boundary conditions on the gravitational perturbations at the asymptotic past and future. As such, the in-out construction is appropriate for calculating instantaneous conservative forces on the compact objects and for computing the instantaneous total power emitted in gravitational waves (at least, in a particular method), among other things. 

One of the goals of this paper is to emphasize that the symmetric nature of the Feynman propagator, which is a consequence of the scattering boundary conditions employed in the in-out formalism, implies that it is unsuitable for self-consistently and systematically describing time-asymmetric processes related to dissipation and radiation reaction in compact binaries \footnote{To be clear, the authors of \cite{GoldbergerRothstein:PRD73} never claimed that the in-out path integral would describe these correctly.}. We stress that this is {\it not} a systematic flaw or shortcoming of NRGR but instead arises from not imposing retarded boundary conditions on the radiated gravitational perturbations.

To address these and other issues in NRGR here we instead
implement the ``in-in" formalism to enforce {\it retarded} boundary conditions in a path integral framework. The in-in construction is an initial value formulation that evolves the system in real time from a given initial state and allows for the final state to be determined dynamically given only initial data. This is in contrast to the {\it ab initio} stipulation of the final state in the in-out construction for scattering processes. 

The in-in approach was first introduced by Schwinger \cite{Schwinger:JMathPhys2} as a way of computing expectation values in quantum mechanics from a path integral formalism and was further developed by others in \cite{BakshiMahanthappa:JMathPhys4_1, BakshiMahanthappa:JMathPhys4_2, Keldysh:JEPT20, ChouSuHao:PRB22, ChouZuHaoYu:PhysRep118, SuChenYuChou:PRB37, DeWitt1986, Jordan:PRD33, CalzettaHu:PRD35, CalzettaHu:PRD37}.

Since its introduction, the in-in formalism has been extensively applied to problems where an initial value formulation is crucial for describing a system's dynamical evolution, typically involving nonequilibrium processes, from an initial state to an unknown final state. These include semi-classical gravity and stochastic gravity (see \cite{HuVerdaguer:LRR} and references therein), inflationary cosmology, quark-gluon plasmas, disoriented chiral condensates, thermal field theory, Bose-Einstein condensates and quantum Brownian motion, to name a few. See \cite{CalzettaHu} for corresponding references. The in-in formalism is also useful for addressing issues related to the quantum-to-classical transition (e.g., decoherence), macroscopic coherence, quantum kinetic theory, noise and fluctutations in open quantum systems, among other things \cite{Weiss, CalzettaHu}.

In the extreme mass ratio inspiral scenario, the in-in formalism is crucial to guarantee the causal evolution of the binary in a curved background spacetime (e.g., Kerr) and has been successfully used in \cite{Galley:EFT1} to rederive the first order self-force \cite{MinoSasakiTanaka:PRD55, QuinnWald:PRD56} acting on the small compact object.

In this paper the in-in framework is used to derive, in the context of NRGR, the well-known compact object equations of motion with radiation reaction at 2.5PN order (first derived by Burke and Thorne \cite{BurkeThorne:Relativity, Thorne:AstrophysJ158, Burke:JMathPhys12}) and the quadrupole gravitational waves emitted by the binary. The in-in formulation of NRGR should also be useful for deriving real-time quantities and the hereditary terms  (e.g., memory and tail integrals) that appear in higher order expressions for the metric components (the gravitational waveform) and the radiated power.

In Section \ref{sec:radEFT} we provide a brief description of the in-out formulation of the radiation sector in NRGR. In Section \ref{sec:qft} we provide a pedagogical presentation of the in-out framework and discuss its shortcomings for describing the real-time causal propagation of gravitational waves. We also review the in-in formalism. We then apply the in-in approach to derive the well-known results for the 2.5PN radiation reaction forces and for the emitted quadrupole gravitational radiation in Section \ref{sec:ctp}. In Appendix \ref{sec:cleft}, we apply the in-in framework to formulate the equivalent classical effective field theory (ClEFT) approach of Kol and Smolkin \cite{KolSmolkin:PRD77} in a form suitable to self-consistently derive radiation reaction and other real-time quantities. This provides an alternative derivation of the in-in approach to NRGR. In this paper we focus on non-spinning compact objects and use the same conventions as \cite{GoldbergerRothstein:PRD73}.

\section{The in-out formulation of NRGR}
\label{sec:radEFT}

The central quantity in the NRGR paradigm is the effective action, $S_{eff}$. At the orbital scale $r$ of the binary, the compact objects can effectively be treated as point particles 
 interacting with nearly instantaneous potentials $H_{\mu\nu}$ and coupled to long wavelength ($\lambda \gg r$), slowly varying, external radiation fields $\barh_{\mu\nu}$. In the ``in-out" path integral formulation 
the effective action is given by
\begin{eqnarray}
	e^{i S_{eff} } = \int {\cal D}\barh_{\mu\nu} \int {\cal D} H_{\mu\nu} \, \exp \Bigg\{i S [ \eta + \barh + H] + i  \sum_{K=1}^2 S_{pp} [ x_K(t), \eta+\barh+H] \Bigg\}
\end{eqnarray}
where $S$ is the (gauge-fixed) Einstein-Hilbert action, $S_{pp}$ is the point particle action for each compact object in the binary and the index $K$ labels each particle. 

Integrating out the potential gravitons $H_{\mu\nu}$ from the theory at the orbital scale (in the Lorenz gauge and on the long wavelength background spacetime) schematically gives for the effective action
\begin{eqnarray}
	e^{i S_{eff} } = \int {\cal D} \barh_{\mu\nu} \, \exp \Bigg\{ i S [ \eta + \barh] + \parbox{16mm}{\includegraphics[width=16mm]{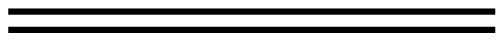}} + \parbox{16mm}{\includegraphics[width=16mm]{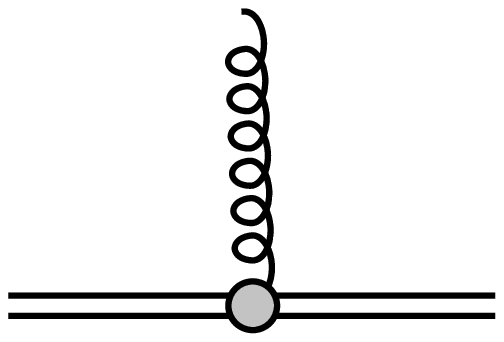}} + \parbox{16mm}{\includegraphics[width=16mm]{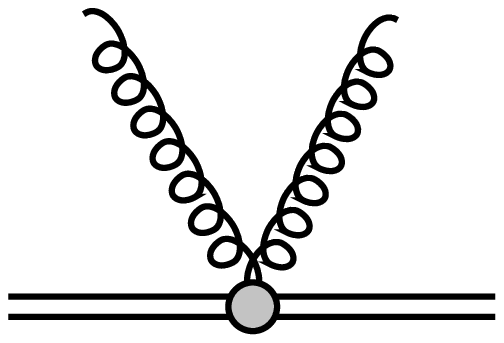}} + \cdots \Bigg\}  \label{effaction1}
\end{eqnarray}
where a curly line denotes a radiation graviton and the double solid line represents the compact binary. We take the radiation graviton to be in the Lorenz gauge
\begin{eqnarray}
	\partial^\alpha \barh_{\alpha \beta} = \frac{1}{2} \partial_\beta \barh^\alpha _{~\alpha}
\end{eqnarray}
throughout the remainder. The diagrams in (\ref{effaction1}) can be written perturbatively in powers of the relative velocity $v$ of the binary by matching onto the effective theory at the orbital scale so that
\begin{equation}
\begin{array}{cccccccc}
	\parbox{16mm}{\includegraphics[width=16mm]{conservative}} & = & \parbox{16mm}{\includegraphics[width=16mm]{conservative}} & + & \parbox{16mm}{\includegraphics[width=16mm]{conservative}}  & + & \parbox{16mm}{\includegraphics[width=16mm]{conservative}} & + \cdots \equiv i S_{v^0 L^1} + i S_{v^2 L^1} + i S_{v^4 L^1} + \cdots \\
	& & v^0 & & v^2 & & v^4 &  
	\label{conservative0}
\end{array} .
\end{equation}
Here $L \sim mv r$ is the typical angular momentum of the binary.
The first diagram in (\ref{conservative0}) represents the leading order Newtonian contribution to the effective action,
\begin{eqnarray}
	i S _{v^0 L^1} = i \int dt \Bigg\{ \sum_K \frac{1}{2} m_K \bv^2_K + \frac{ G_N m_1 m_2 }{ | \bx_1 - \bx_2 |^2 } \Bigg\}  , \label{newton0}
\end{eqnarray}
the second diagram provides the 1PN corrections \cite{GoldbergerRothstein:PRD73} (from which the Einstein-Infeld-Hoffman equations of motion \cite{EinsteinInfeldHoffmann:AnnMath39} are derived), and the third diagram gives the 2PN corrections, which have been recently computed using the NRGR approach in \cite{GilmoreRoss:0810.1328} and agree with previous calculations using traditional methods (see \cite{Blanchet:LRR}).

The diagrams in (\ref{conservative0}) do not couple to radiation gravitons, implying that there is no mechanism for the compact objects to dissipate energy (aside from possible tidal heating in neutron stars at higher orders) via radiation reaction. However, the remaining diagrams in (\ref{effaction1}) describe the interactions of the compact binary with radiation gravitons and are responsible for driving the inspiral toward coalescence from the emission of gravitational waves.

The diagrams containing a radiation graviton in (\ref{effaction1}) represent a multipole expansion in terms of the size of the binary $r$ and the wavelength of the radiation $\lambda$. These lengths are related, since $r / \lambda \sim v$, which
implies that these diagrams scale as powers of $v$. For example, the second diagram in (\ref{effaction1}) represents the full coupling of a single radiation graviton to the binary's center of mass and has a multipole expansion given schematically by
\begin{equation}
\begin{array}{cccccccc}
	\parbox{16mm}{\includegraphics[width=16mm]{radiation1}} & = & \parbox{16mm}{\includegraphics[width=16mm]{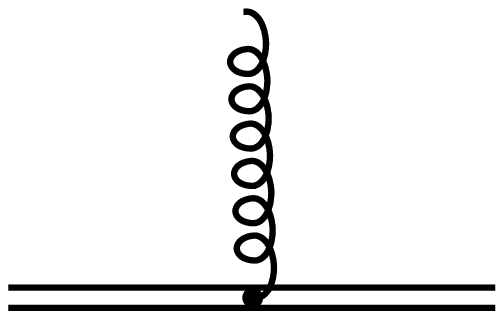}} & + & \parbox{16mm}{\includegraphics[width=16mm]{radiation1com}}  & + & \parbox{16mm}{\includegraphics[width=16mm]{radiation1com}} & + \cdots \\
	& & v^{1/2} & & v^{3/2} & & v^{5/2} &
	\label{radiation0}
\end{array}
	\equiv i S_{v^{1/2} L^{1/2}} + i S_{v^{3/2} L^{1/2}} + i S_{v^{5/2} L^{1/2}} + \cdots
\end{equation}
where the first three diagrams above are computed explicitly in \cite{GoldbergerRothstein:PRD73}, 
\begin{eqnarray}
	i S_{v^{1/2} L^{1/2} } &=& - \frac{i}{2 m_{pl} } \int dt \sum_K m_K \barh_{00} (t, \bzero) \\
	i S_{v^{3/2} L^{1/2} } &=& - \frac{ i }{ m_{pl}} \int dt \sum_K m_K x_K^i \barh_{i0} (t, \bzero) = 0 {\rm ~~in ~center ~of ~mass ~frame} \label{dipole1} \\
	i S_{v^{5/2} L^{1/2} } &=& \frac{ i }{2 m_{pl} } \int dt \Big\{ - E \, \barh_{00} (t, \bzero) - \epsilon_{ijk} L_k \partial_j \barh_{i0} (t, \bzero) + Q_{ij} (t) R_{0i0j} (t, \bzero) \Big\}  . \label{quad1}
\end{eqnarray}
and their scaling with $(v, L)$ is given in the subscript.
Here $E$ is the leading order (Newtonian) total energy, ${\bf L}$ is the leading order angular momentum, $Q_{ij}$ is the quadrupole moment of the binary
\begin{eqnarray}
	Q^{ij} (t) = \sum_{K=1}^2 m_K \left[ x_K^i (t) x_K^j (t) - \frac{ 1}{3} \delta^{ij} \bx_K^2 (t) \right]
\end{eqnarray}
and $R_{0i0j}$ is the linearized Riemann tensor, given by
\begin{eqnarray}
	R_{0i0j} = \frac{ 1}{2} \Big( \partial_0^2 \barh_{ij} + \partial_i \partial_j \barh_{00} - \partial_0 \partial_i \barh_{j0} - \partial_0 \partial _j \barh_{i0} \Big) + O(\barh^2) .  \label{linearRiem1}
\end{eqnarray}
In obtaining these expressions, the radiation graviton field $\barh_{\mu\nu} (t, \bx)$ is expanded in multipoles around the center of mass, which is taken to be the origin of coordinates. Explicitly writing the velocity expansion into the effective action gives
\begin{eqnarray}
	e^{i S_{eff} } = \int {\cal D} \barh_{\mu\nu} \, \exp \Big\{ i S_{pot} [ \bx_1, \bx_2, \barh]  \Big\}  .  \label{effactionrad0}
\end{eqnarray}
where
\begin{eqnarray}
	S_{pot} [ \bx_1, \bx_2, \barh] = S [ \eta + \barh ] + S_{v^0 L^1} + S_{v^2 L^1} + S_{v^4 L^1 } + \cdots + S_{v^{1/2} L^{1/2} } + S_{v^{3/2} L^{1/2} } + S_{v^{5/2} L^{1/2} } + \cdots 
	\label{poteffaction0}
\end{eqnarray}
is the (effective) action obtained by integrating out the potential gravitons.
This expression for the in-out effective action is useful for deriving the conservative forces acting on the compact objects \cite{GoldbergerRothstein:PRD73, GilmoreRoss:0810.1328}, for computing the instantaneous power emitted in gravitational waves \cite{GoldbergerRothstein:PRD73, CardosoDiasFigueras:PRD78}, and can even be used to recover the Schwarzschild metric of a single compact object \footnote{Private communication with Ira Rothstein.}. However, to derive the radiation reaction forces in the equations of motion and to compute the multipole expansion of the emitted gravitational waves in NRGR requires imposing retarded boundary conditions by using an initial value (or ``in-in") formulation to describe these quantities in real-time. To demonstrate that the in-out path integral formulation above is inadequate for these latter purposes, let us attempt to derive the leading order radiation reaction forces acting on the compact objects using (\ref{effactionrad0}).

\subsection{Radiation reaction}

\begin{figure}
	\center
	\includegraphics[width=10.5cm]{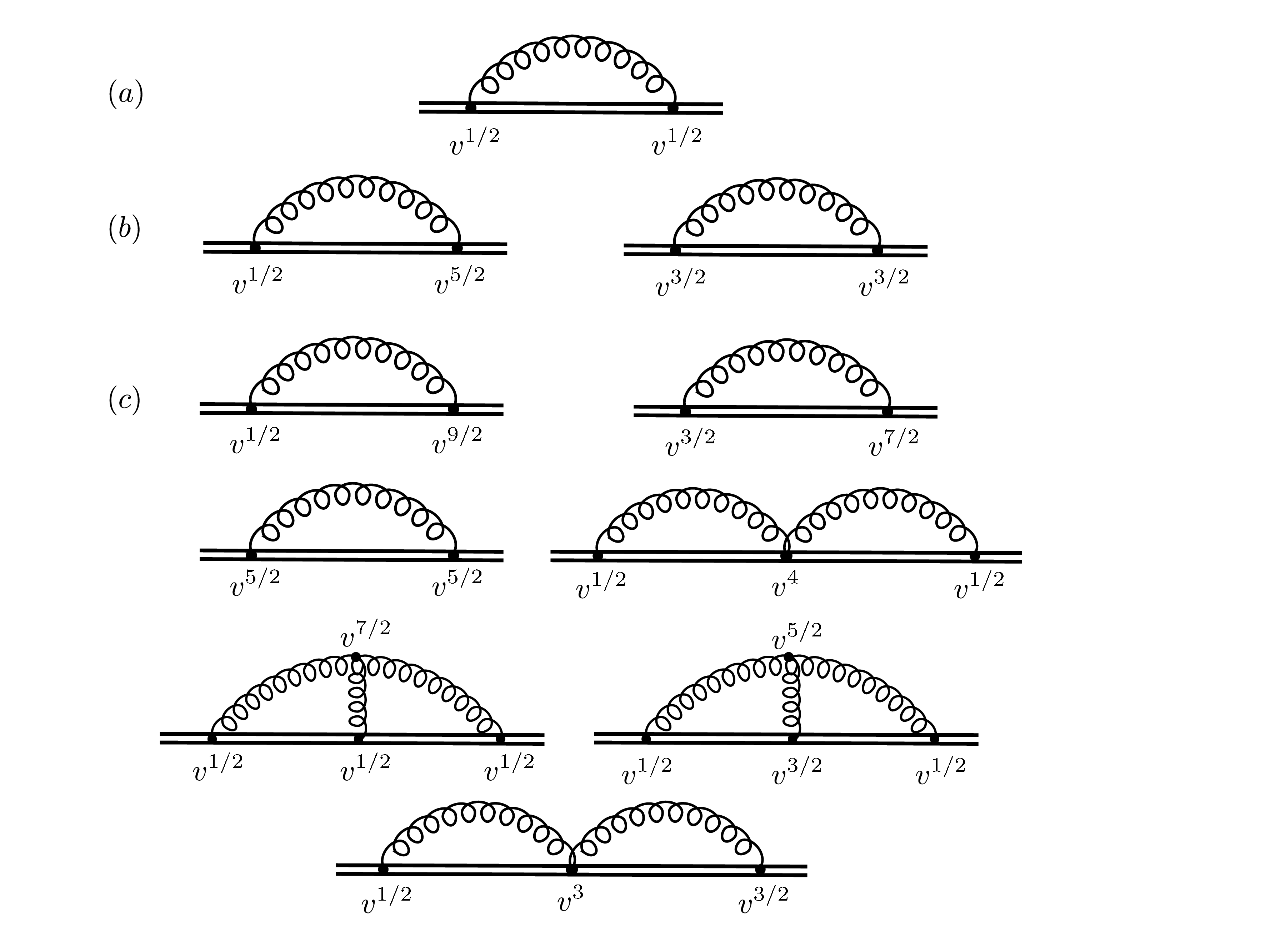}
	\caption{The diagrams that potentially contribute to radiation reaction at (a) 0.5PN, (b) 1.5PN and (c) 2.5PN orders.}
	\label{fig:rr1}
\end{figure}

Integrating out the radiation gravitons involves calculating all tree-level and connected Feynman diagrams to the desired order in $v$ and is equivalent to performing the path integral over $\barh_{\mu\nu}$. The resulting theory describes the motion of the compact objects (effectively as point particles)  subject to radiation reaction due to the backreaction of gravitational wave emission. The diagrams that can contribute to radiation reaction through 2.5PN are given in Fig.(\ref{fig:rr1}). 

In Appendix \ref{app:zerodiagram} we show that diagrams with a time-dependent (but otherwise generic) vertex coupled via a radiation graviton to a {\it conserved} quantity will not contribute to the equations of motion or the radiated power. The vertex associated with $S_{v^{1/2} L^{1/2}}$ is proportional to the total mass of the binary $M$, which is conserved at leading order, implying that the single 0.5PN diagram in Fig.(\ref{fig:rr1}a) vanishes. Likewise, the $v^{3/2}$ vertex is zero in the center of mass frame; see (\ref{dipole1}). Therefore, both diagrams in Fig.(\ref{fig:rr1}b) vanish. Similar reasoning implies that only the diagram containing two $v^{5/2}$ vertices in Fig.(\ref{fig:rr1}c) does not automatically equal zero. Furthermore, the first two terms in (\ref{quad1}) are conserved at this order suggesting that these give a vanishing contribution in the effective action at 2.5PN. Indeed, only the quadrupole term in (\ref{quad1}) is relevant so that
\begin{eqnarray}
	{\rm Fig.(\ref{fig:rr1}c)} = i S_{eff} ^{2.5{\rm PN}} = \left( \frac{1}{2} \right) \left( \frac{i}{2 m_{pl}} \right)^2 \int dt \int dt^\prime Q_{ij}(t) \big\langle R_{0i0j} (t, \bzero) R_{0k0l} (t^\prime, \bzero) \big\rangle Q_{kl}(t^\prime)  \label{inouteffaxnrr0}
\end{eqnarray}
The leading factor of $1/2$ is the symmetry factor of the diagram. The imaginary part of Fig.(\ref{fig:rr1}c) is originally computed in \cite{GoldbergerRothstein:PRD73} and generalized to $d$ spacetime dimensions in \cite{CardosoDiasFigueras:PRD78}. The real part of this diagram yields a contribution to the equations of motion upon variation with respect to the particle coordinates.

The two-point function of the product of linearized Riemann tensors is 
\begin{eqnarray}
	\big\langle R_{0i0j} (t, \bzero) R_{0k0l} (t^\prime, \bzero) \big\rangle = \frac{i}{20} \int_{k^0, \bk} e^{-i k^0 (t-t^\prime) } \frac{ (k^0)^4 }{ (k^0)^2 - \bk^2 + i \epsilon } \left[ \delta_{ik} \delta_{jl} + \delta_{ik} \delta _{jl} - \frac{2}{3} \delta_{ij} \delta_{kl} \right]  \label{prodRiem1} 
\end{eqnarray}
so that
\begin{eqnarray}
	 i S_{eff} ^{2.5{\rm PN}} = - \frac{ i }{ 80 m_{pl}^2} \int  dt \int dt^\prime \, Q_{ij} (t) Q_{ij}(t^\prime) \int _{k^0, \bk} e^{-i k ^0 (t-t^\prime) } \frac{ (k^0)^4 }{ (k^0)^2 - \bk^2 + i \epsilon}  \label{radrxn0}
\end{eqnarray}
and the distributional identity in (\ref{PV0}) implies
\begin{eqnarray}
	i S_{eff}^{2.5{\rm PN} } &=& - \frac{ 2 \pi i G_N}{ 5} \int dt \int dt^\prime Q_{ij}(t) Q_{ij}(t^\prime) \int_{\bk} \Bigg\{ PV \int_{k^0} \, \frac{ (k^0)^4 e^{- i k^0 (t-t^\prime) } }{ (k^0)^2 - \bk^2 } + i \pi \int _{k^0} (k^0)^4 e^{-i k^0 (t-t^\prime) } \delta \big( (k^0)^2 - \bk^2 \big) \Bigg\}  .  \nonumber \\
	&&
\end{eqnarray}
In the $t^\prime$ integral of the first term, $t$ is constant so we may shift the variable to $s = t^\prime - t$ and expand $Q_{ij} (t+s)$ in powers of $s$, giving
\begin{eqnarray}
	i S_{eff} ^{2.5{\rm PN} } &=& - \frac{ 2 \pi i G_N }{ 5 } \int dt \, Q_{ij} (t) \sum_{n=0}^\infty \frac{ 1}{ n!} \frac{ d^n}{dt^n} Q_{ij} (t) \int ds \int _{\bk} PV \int_{k^0} \frac{ s^n (k^0)^4 e^{i k^0 s} }{ (k^0)^2 - \bk^2 } \nonumber \\
	&& + \frac{ 2 \pi^2 G_N }{5 } \int_{\bk, k^0} (k^0)^4 \delta \big( (k^0)^2 - \bk^2 \big) \, \tilde{Q}_{ij} (k^0) \tilde{Q}_{ij} (- k^0)  
\end{eqnarray}
where $\tilde{Q}_{ij} (k^0)$ is the Fourier transform of the binary's quadrupole moment; see (\ref{fourier1}) for our conventions. In the first term, we recognize the master integral $I(n, 4, 0)$ defined in (\ref{master0}) and evaluated in Appendix \ref{app:master}. This implies that only the $n=5$ term 
contributes, so that
\begin{eqnarray}
	i S_{eff}^{2.5{\rm PN} } = - \frac{ i G_N }{ 10} \int dt \, Q_{ij} (t) \frac{ d^5 Q_{ij}(t) }{ dt^5} + \frac{ \pi G_N}{ 5} \int \frac{ d^3 k}{ (2\pi)^3} \, | \bk |^3 \big| \tilde{Q}_{ij} ( | \bk | ) \big| ^2  \label{radrxn1}
\end{eqnarray}
with $I(5,4,0) = 5! / (4\pi)$ from (\ref{master1}).

Our expression for the imaginary part of the effective action in (\ref{radrxn1}) agrees with \cite{GoldbergerRothstein:PRD73, CardosoDiasFigueras:PRD78} and yields the quadrupole formula for power loss upon inserting another factor of $|\bk|$ to the integrand. The real part is proportional to the time integral of the Burke-Thorne ``reaction potential" \cite{BurkeThorne:Relativity, Thorne:AstrophysJ158,  Burke:JMathPhys12}
\begin{eqnarray}
	\Phi _{BT} (t, \bx) = \frac{ G_N}{5} x^i x^j \frac{ d^5 Q_{ij}(t) }{ dt^5}  \label{burkethorne0}
\end{eqnarray}
evaluated at the positions of the two particles. However, two integration by parts in the first term of (\ref{radrxn1}) reveals that
\begin{eqnarray}
	{\rm Re} \, S_{eff} ^{2.5{\rm PN}} = - \frac{ G_N }{ 10 } \int dt \, \ddot{Q}_{ij} (t) \dddot{Q}_{ij} (t) = - \frac{ G_N}{ 20} \int dt \, \frac{ d}{ dt} \big( \ddot{Q}_{ij} \big)^2, \label{radrxn2}
\end{eqnarray}
which is a total time derivative and does not yield a contribution to the equations of motion. 

In the next section, we compute the emitted gravitational waves in the quadrupole approximation using the in-out effective action (\ref{effactionrad0}) and demonstrate that the gravitational radiation does not propagate causally from the source.

\subsection{Quadrupole radiation}
\label{sec:gwinout}

The first few diagrams contributing to the gravitational radiation emitted by the compact binary are shown in (\ref{radiation0}). In the physically relevant transverse-traceless (TT) gauge, the first two diagrams in (\ref{radiation0}) vanish. This is expected since these correspond to monopole and dipole radiation, which are pure gauge effects and can be transformed away. This leaves the third diagram in (\ref{radiation0}), wherein the first two terms of (\ref{quad1}) do not contribute in the TT gauge. Applying the Feynman rules gives the leading order contribution
\begin{eqnarray}
	\parbox{16mm}{\includegraphics[width=16mm]{radiation1}} = \barh_{ij} ^{TT} (t, \bx) = \frac{ i }{ 2m_{pl} } \Lambda_{ij,kl} (\hat{\bn}) \int dt^\prime \, \big\langle \barh_{kl} (t, \bx) R_{0m0n} (t^\prime, \bzero) \big\rangle Q_{mn} (t^\prime)   \label{gw0}
\end{eqnarray}
where $\hat{\bn}$ is the unit vector pointing in the direction of propagation and $\Lambda_{ij,kl}$ is given explicitly by
\begin{eqnarray}
	\Lambda _{ij,kl} (\hat{\bn} ) = \delta_{ik} \delta_{jl} - \frac{1}{2} \delta_{ij} \delta _{kl} - n_j n_l \delta_{ik} - n_i n_k \delta_{jl} + \frac{1}{2} n_k n_l \delta_{ij} + \frac{1}{2} n_i n_j \delta_{kl} + \frac{1}{2} n_i n_j n_k n_l  ,
\end{eqnarray}
which is pair-wise symmetric in $ij, kl$. See \cite{Maggiore} for a more detailed discussion regarding the relationship between the Lorenz and TT gauges.

From the expression for the linearized Riemann tensor in (\ref{linearRiem1}) we see that the two-point function in (\ref{gw0}) is
\begin{eqnarray}
	\big\langle \barh_{kl} (t, \bx) R_{0m0n} (t^\prime, \bzero) \big\rangle = \frac{1}{2} \Big\langle \barh_{kl}(t,\bx) \Big[ \partial_{t^\prime}^2 \barh_{mn} (t^\prime, \bzero) + \partial_m \partial_n \barh_{00} (t^\prime, \bzero) - \partial_{t^\prime} \partial_m \barh_{n0} (t^\prime, \bzero) - \partial_{t^\prime} \partial_n \barh_{m 0} (t^\prime, \bzero) \Big] \Big\rangle  .
\end{eqnarray}
Only the first term contributes, so that
\begin{eqnarray}
	m_{pl}^{-1} \barh_{ij}^{TT} (t, \bx) = \frac{ i }{ 4 m_{pl}^2} \Lambda_{ij,kl} (\hat{\bn} ) \int dt^\prime \, D^F_{klmn} (t-t^\prime, \bx) \ddot{Q}_{mn} (t^\prime)  \label{inoutquad0}
\end{eqnarray}
where we have integrated by parts twice to move the time derivatives onto the quadrupole moment. 

Taking the (physically relevant) real part of $\barh_{ij}^{TT}$ using
\begin{eqnarray}
	{\rm Re} \, i D^F_{klmn} (t-t^\prime, \bx) = \frac{1}{2} \Big( D^{ret}_{klmn} + D^{adv} _{klmn} \Big) = - \frac{1}{4\pi} \delta \left( (t - t^\prime)^2 - \bx^2 \right) P_{klmn}  \label{gravfeynman0}
\end{eqnarray}
where $P_{klmn} = ( \eta_{km} \eta_{ln} + \eta_{kn} \eta_{lm} - \eta_{kl} \eta_{mn} ) / 2$ and performing the time integral gives
\begin{eqnarray}
	m_{pl}^{-1} {\rm Re} \, \barh_{ij}^{TT} (t, \bx ) = - \frac{ G_N }{ | \bx | } \Lambda_{ij, kl} (\hat{\bn} ) \Big[ \ddot{Q}_{kl} (t - | \bx |) + \ddot{Q}_{kl} (t + | \bx |) \Big]  . \label{inoutquad1}
\end{eqnarray}
We see that the gravitational waves do not causally propagate from the source because of the dependence on the quadrupole moment at the {\it advanced} time $t + | \bx |$. Furthermore, (\ref{inoutquad1}) implies that there is no power emitted by the compact binary. 

To see this, recall that the power emitted in gravitational waves is the surface integral of the radial flux of gravitational waves $t^{0r}$ \cite{Maggiore}
\begin{eqnarray}
	P = \int dA \, t^{0r} = - \int dA \, \langle \partial_0 \barh_{ij} ^{TT} \partial_r \barh_{ij}^{TT} \rangle  \label{power0}
\end{eqnarray}
where angled brackets here denote averaging over many cycles of the gravitational wave and the surface integral is over a very large sphere centered on the binary's center of mass. Using (\ref{inoutquad1}) and the relation $\partial_r f(t \pm |\bx|) = \pm \partial_0 f( t\pm |\bx| )$ one finds that
\begin{eqnarray}
	t^{0r} \propto \big\langle  \dddot{Q}_{ij}(t- |\bx| ) \dddot{Q}_{ij} (t-|\bx|) \big\rangle - \big\langle \dddot{Q}_{ij} (t+|\bx|) \dddot{Q}_{ij} (t+|\bx||) \big\rangle
\end{eqnarray}
Both terms can be shown to have the same Fourier transform implying that the flux of purely retarded radiation is cancelled by the flux of purely advanced radiation leading to $t^{0r}=0$ and vanishing power loss. 
This explains the lack of radiation reaction in the equations of motion seen in the previous section.

This raises an interesting observation about the in-out approach. Computing the power loss directly from the gravitational perturbations $\barh_{\mu\nu}$ gives a vanishing result because the scattering boundary conditions imply that there is no net flux of gravitational waves leaving the compact binary in an irreversible and asymmetric manner. However, since the imaginary part of the effective action $S_{eff}^{2.5{\rm PN}}$ is related to decay processes then a non-zero expression for the power loss can be computed in the manner discussed in \cite{GoldbergerRothstein:PRD73}. In particular, this quantity is related to the total number of gravitons emitted by the system during a large time $T\to \infty$
\begin{eqnarray}
	\frac{1}{T} {\rm Im} S_{eff} = \frac{1}{2} \int d \omega d\Omega \frac{ d^2 \Gamma}{d \omega d\Omega}
\end{eqnarray}
where $d\Gamma$ is the differential rate for graviton emission. Therefore, the differential power is $dP = \omega d\Gamma$ so that the integrated power spectrum is
\begin{eqnarray}
	P = \int d\omega d\Omega \, \omega \frac{ d^2 \Gamma}{ d \omega d\Omega}
\end{eqnarray}
and was used to derive the famous quadrupole power loss formula in NRGR in \cite{GoldbergerRothstein:PRD73, CardosoDiasFigueras:PRD78}.

\subsection{Radiation reaction, self-consistency and hereditary effects}
\label{sec:other}

Before continuing, we mention some technical subtleties associated with the in-out calculation of the radiation reaction presented above. There we demonstrated that the initially divergent effective action in (\ref{inouteffaxnrr0}) has a finite real part that is the integral of a time derivative, which implies that the radiation reaction force is zero when derived from a variational principle; see (\ref{radrxn2}). However, the correct expression could be calculated by first varying the divergent effective action (\ref{inouteffaxnrr0}) and then regularizing the divergence in the resulting force \footnote{This method is adopted in \cite{Leibovich:NRGRtalk}.}. It is straightforward to show that doing so yields,
\begin{eqnarray}
	\frac{ \delta {\rm Re} S_{eff}^{2.5 {\rm PN} } }{ \delta x^i _K (t) } = - \frac{ 2 m_K G_N }{ 5} \, \frac{d ^5 Q_{ij} (t) }{ dt^5}    \, x_K^j (t) ,  \label{inoutburke0}
\end{eqnarray}
which is the correct radiation reaction force first given by Burke and Thorne \cite{BurkeThorne:Relativity, Thorne:AstrophysJ158,  Burke:JMathPhys12}. However, we remark that this procedure (of regularizing the variation of the action instead of the action itself)
can lead to inconsistencies within the in-out formalism and to incorrect expressions for history-dependent or hereditary quantities.

For example, the finite part of the real part of the effective action in (\ref{inouteffaxnrr0}) gives the integral of a total time derivative (\ref{radrxn2}) and yields a radiation reaction force equal to zero and not (\ref{inoutburke0}). Either one has a regular effective action that yields a vanishing radiation reaction or one has a divergent and ill-defined Lagrangian that provides a finite and correct expression for the radiation reaction force (at least, if it is local in time). Therefore, it seems that the in-out formalism does not give a finite effective action that {\it simultaneously} provides the correct expression for the radiation reaction force.

In addition, the force (\ref{inoutburke0}) is not conservative and represents a dissipative component to the compact object equations of motion that should correspond with the emission of gravitational waves satisfying retarded boundary conditions. However, the gravitational perturbations are still given by (\ref{inoutquad1}) and yield no power loss (when computed from the field directly as in (\ref{power0}) instead of from the imaginary part of the effective action) despite the appearance of physical radiation reaction forces on the compact objects according to (\ref{inoutburke0}). 

Finally, consider the extreme mass ratio inspiral (see \cite{Poisson:LRR} for a comprehensive pedagogical review). In this system, the radiation reaction (more appropriately called ``self-force") is a force on the small compact object that depends on the entire past history of the small compact object's worldline and on the configuration of the gravitational perturbations it couples to in the past. As such, the self-force at proper time $\tau$ is affected only by gravitational perturbations that have evolved from within the past null cone of the worldline. It is straightforward to show that using the in-out formalism to calculate the self-force by first varying the effective action and then regularizing the resulting self-force leads to an extra dependence 
from the {\it future} null cone due to the advanced propagator, thus violating causality. 

In general, the in-out formalism may not give correct expressions for quantities that ought to have a causal history-dependence. This likely includes the hereditary terms in the PN expansion of the near-zone metric, in the radiative field at future null infinity, and in the radiated power at higher orders (1.5PN relative order and beyond). If one wants to use NRGR to systematically and self-consistently compute the kinds of quantities discussed throughout this section then one should implement a framework different from the in-out prescription discussed here.

In the next section, after presenting the underlying assumptions and construction of the in-out framework, we discuss the ``in-in" approach, which provides a self-consistent formalism suitable for describing the causal evolution of the compact binary as it undergoes the real-time processes of dissipation and backreaction from the emission of gravitational radiation that causally propagates to future null infinity.

\section{Comparing the In-out and in-in formalisms}
\label{sec:qft}

We strongly emphasize that the problems encountered in the previous section do {\it not} stem from a systematic flaw with NRGR. Rather, these issues follow from using a particular formalism for computing the effective action that does not implement the (asymmetric) retarded boundary condition on radiation gravitons. This condition is necessary for self-consistently describing the real-time processes of dissipation (via radiation reaction) and backreaction (from gravitational wave emission) that occur during the dynamical evolution of the compact binary. 
In this section, we provide motivation for these statements and review the framework appropriate for addressing these issues. Our discussion will be somewhat pedagogical. 
The reader already familiar with the in-out and in-in formalisms can proceed ahead to Section \ref{sec:ctp} for the in-in representation of NRGR or to Section \ref{sec:feynrules} for the corresponding Feynman rules.

We first present the in-out formulation of NRGR within the more standard notation of quantum field theory found in many common texts (e.g., \cite{Peskin, Ryder, Itzykson}). This will allow for a more transparent presentation of the details regarding the in-out and in-in formalisms. 

In the in-out path integral formulation (i.e., the one commonly encountered in textbooks), the generator of connected correlation functions $W$ is a functional of several auxiliary sources that couple to the radiation gravitons and to the coordinates of each particle's trajectory:
\begin{eqnarray}
	e^{i W [ \bj_1, \bj_2, J^{\mu\nu} ] } = \int \prod_{K=1}^2 {\cal D} \bx_K \int {\cal D} \barh_{\mu\nu} \, \exp \Bigg\{ i S_{{pot}} [ \bx_1, \bx_2, \barh ]  + i \sum_{K=1}^2 \int dt \, \bj_K \cdot \bx_K + i \int d^4 x \, J^{\mu\nu} \barh_{\mu\nu} \Bigg\},  \label{inouteffaxn0}
\end{eqnarray}
where $S_{{pot}}$ is the (effective) action describing the two compact objects coupled to the radiation gravitons, which is computed by integrating out the potential gravitons from the theory \cite{GoldbergerRothstein:PRD73}; see (\ref{poteffaction0}).
Variation of $W$ with respect to $J^{\mu\nu}$ gives the one-point function of the radiation graviton (i.e., the quadrupole radiation computed in Section \ref{sec:gwinout})
\begin{eqnarray}
	\big\langle \hat{\barh}_{\mu\nu} (x^\alpha) \big\rangle_{{\rm in-out}} = \frac{ \delta W}{ \delta J^{\mu\nu} (x^\alpha) } \Bigg|_{\bj_1 = \bj_2 = J^{\mu\nu}=0}  \label{pneftgw0}
\end{eqnarray}
while a similar variation with respect to $\bj_K(t)$ gives $\langle \hat{\bx}_K (t) \rangle_{\rm in-out}$. 
A partial Legendre transform of $W$
\begin{eqnarray}
	\Gamma [ \langle \hat{\bx}_1 \rangle , \langle \hat{\bx}_2 \rangle, J^{\mu\nu} ] = W [ \bj_1, \bj_2, J^{\mu\nu} ] - \sum_{K=1}^2 \int dt \, \bj_K \cdot \langle \hat{\bx}_K \rangle _{{\rm in-out}}
\end{eqnarray}
gives the functional whose variation of the real part yields the equations of motion for the orbital motion of the compact objects
\begin{eqnarray}
	0 = \frac{  \delta {\rm Re} \Gamma }{ \delta \langle \hat{\bx}_K (t) \rangle_{{\rm in-out}} } \Bigg|_{\bj_1 = \bj_2 = J^{\mu\nu} = 0}  .
\end{eqnarray}
These equations describe the evolution of $\langle \hat{\bx}_K (t) \rangle_{{\rm in-out}}$, which are assumed to be the classical variables $\bx_K(t)$ (since the masses of the compact objects are astronomically large implying that there are negligible quantum fluctuations of $\hat{\bx}_K(t)$). 

Actually, (\ref{inouteffaxn0}) describes a first principles description of the compact binary where the trajectories of the compact objects are quantum mechanical and the gravitons are quantum fields. As such, a mechanism for decohering the quantum trajectories $\hat{\bx}_K (t)$ is needed to ensure that the compact objects evolve within a (semi-)classical limit. However, this requires a more detailed discussion of the suppression of quantum interference effects, which is usually given in a density matrix formulation and is intrinsically an initial value problem. 
See \cite{JohnsonHu:PRD65, GalleyHu:PRD72, GalleyHuLin:PRD74, CalzettaHu} and references therein for further discussion on this and related points.

Nevertheless, in the assumed classical limit, the path integral over the compact object trajectories is dropped and $\langle \hat{\bx}_K(t) \rangle_{{\rm in-out}}$ is taken as an external field from the point of view of the radiation gravitons. As we are about to show below, $\langle \hat{\bx}_K(t) \rangle_{{\rm in-out}}$ (as derived from $W$ above) is not the expectation value of the position operator. Therefore, interpreting the (in-out) one-point function as a classical variable, which ought to be selected by physical mechanisms of decoherence, is inaccurate and can lead to certain problems when trying to describe the classical limit of a quantum theory. Indeed, classical variables are real and evolve causally in time.

\subsection{One-point functions in the in-out formalism}

We demonstrated earlier that using the in-out formalism does not give a contribution to the radiation reaction forces from an action principle and does not describe the causal propagation of gravitational waves emitted from the binary. In this section, we show why this occurs. 

Let us consider a much simpler theory to focus our attention on without having the distraction of cumbersome details that are otherwise irrelevant for this discussion.
Consider a real, massive and linear scalar field in flat spacetime coupled linearly to a physical source $Q(t, \bx)$
\begin{eqnarray}
	S [ \phi]  = \frac{1}{2} \int d^4x \, \big[ \partial_\alpha \phi \partial^\alpha \phi + m^2 \phi^2 + 2 Q  \phi \big]  .  \label{KGaction0}
\end{eqnarray}
The last term is analogous to the interaction terms, $S_{v^{1/2} L^{1/2} } + \cdots$, which are linear in the radiation graviton field and depend on the trajectories of the compact objects.

In the interaction picture, a state $| \alpha \rangle$ evolves from $t = - \infty$ according to $| \alpha ; t \rangle = \hat{U}_{J+Q} (t, -\infty) | \alpha \rangle$ where $J(t, \bx)$ is an auxiliary source whose role will be obvious below. The time-evolution operator is
\begin{eqnarray}
	\hat{U}_{J+Q} (t, t^\prime) \equiv T \, \exp \Bigg\{ i \int_{t^\prime} ^t d\tau \int d^3 x \, \big[ J(\tau, \bx) + Q(\tau, \bx) \big] \hat{\phi}_I (\tau, \bx) \Bigg\}  \label{timeevo0}
\end{eqnarray}
where $T$ is the time-ordering operator and $\hat{\phi}_I$ is the field in the interaction picture, which is related to the Heisenberg picture by
\begin{eqnarray}
	\hat{\phi}_I (t, \bx) = \hat{U}_{J+Q} (t, -\infty) \hat{\phi}_H (t, \bx) \hat{U}_{J+Q} (-\infty, t)  . \label{intrxnHeis0}
\end{eqnarray}

The generating functional $W[J]$ for connected correlation functions of the field operator is defined as \cite{Peskin, BirrellDavies}
\begin{eqnarray}
	e^{ i W [ J ] } \equiv \langle 0, {\rm out} | \hat{U}_{J+Q} (\infty, -\infty) | 0, {\rm in} \rangle .  \label{inout0}
\end{eqnarray}
Here $|0, {\rm in} \rangle$ denotes the vacuum state in the distant past (the ``in" vacuum) and $| 0, {\rm out} \rangle$ is the vacuum state in the distant future (the ``out" vacuum), which does not necessarily equal $|0, {\rm in} \rangle$, especially in a general background spacetime. The presence of these vacua in (\ref{inout0}) provides the name of this formalism, the so-called ``in-out" approach. When the auxiliary source $J$ and the interaction $Q$ are adiabatically turned on and off (as is usually assumed in quantum field theory) we can set the in- and out-vacua equal to the state $| 0 \rangle$ since there is a preferred vacuum selected by the Poincare symmetry of the underlying background Minkowski spacetime. Therefore, we write (\ref{inout0}) as
\begin{eqnarray}
	e^{i W [ J ] } = \langle 0 | \hat{U}_{J+Q} (\infty, -\infty) | 0 \rangle  .  \label{inout1}
\end{eqnarray}
The one-point function of the field is the functional derivative of $W$ with respect to the source so that 
\begin{eqnarray}
	\langle \hat{\phi}_H (t, \bx) \rangle_{{\rm in-out}} \equiv \frac{ \delta W }{ \delta J(t, \bx) } \Bigg|_{J=0}  \label{inoutonept0}
\end{eqnarray}
and is the analogue of the one-point function in (\ref{pneftgw0}). It is commonly implied that (\ref{inoutonept0}) is an expectation value of the field because the in- and out-vacua are equal. We will now show that while the in-out one-point function (as derived from $W$) is the expectation value of some operator, it is not of the field.

To see this, we use (\ref{timeevo0}) to show that
\begin{eqnarray}
	\frac{ \delta \hat{U}_{J+Q} (\infty, -\infty) }{ \delta J (t, \bx) } = i \hat{U}_{J+Q} (\infty, t) \hat{\phi}_I (t, \bx) \hat{U}_{J+Q} (t, -\infty)  \label{timeevo1}
\end{eqnarray}
from which it follows that (\ref{inoutonept0}) can be written as
\begin{eqnarray}
	\langle \hat{\phi}_H (t, \bx) \rangle_{{\rm in-out}} = \langle 0 | \hat{U}_{Q} (+\infty, t) \hat{\phi}_I (t, \bx) \hat{U}_{Q} (t, -\infty) | 0 \rangle .  \label{inoutonept1}
\end{eqnarray}
Writing the right side in terms of the Heisenberg picture field, with $J=0$ in (\ref{intrxnHeis0}), gives
\begin{eqnarray}
	\langle \hat{\phi}_H (t, \bx) \rangle _{{\rm in-out}} &=& \langle 0 | \hat{U}_Q (\infty, -\infty) \hat{\phi}_H (t, \bx) | 0 \rangle  \label{inoutonept12} \\ 
	&=& \langle Q | \hat{\phi}_H (t, \bx) | 0 \rangle  \label{inoutonept2} 
\end{eqnarray}
where $| Q \rangle \equiv \hat{U}_Q (-\infty, \infty) | 0 \rangle$ is the state $| 0 \rangle$ evolved from the distant {\it future} to the distant {\it past} in the presence of the physical interaction $Q$. 
Notice that (\ref{inoutonept12}) is not the expectation value of the field but of the operator $\hat{U}_Q \hat{\phi}_H$, instead. 
Therefore, one-point functions (and, more generally, $n$-point functions) in the in-out formalism are {\it not} true expectation values of the field. 

Instead, (\ref{inoutonept2}) is a matrix element of the field operator for a non-vanishing interaction $Q$ and is generally complex. For our real scalar field it is easy to show that (\ref{inoutonept12}) is not purely real, which also suggests that the in-out one-point function is not a classical field.
(Recall that to make the graviton one-point function purely real, we simply took the real part of the generally complex expression in (\ref{inoutquad0}) using (\ref{gravfeynman0}).)

Most importantly, and not unrelated, is the fact that the in-out one-point function does not describe the retarded propagation of the field perturbations that are induced by $Q$. This can be seen by either computing the right side of (\ref{inoutonept1}) directly or using a path integral representation for $e^{iW}$ and computing the variation as in (\ref{inoutonept0}). Both methods give
\begin{eqnarray}
	\langle \hat{\phi}_H (t, \bx) \rangle _{{\rm in-out}} &=& i \int d^4 y \, D_F (x-y) Q(y)  \\
		&=& \frac{1}{2} \int d^4 y \big[ D_{ret} (x-y) + D_{adv} (x-y) - i D_H (x-y) \big] Q(y)  \label{inoutonept3}
\end{eqnarray}
after using (\ref{feynmanRI}). Note the dependence on the advanced propagator and on the non-local Hadamard two-point function $D_H$, which is non-zero even when $x^\alpha-y^\alpha$ is spacelike. While taking the real part of (\ref{inoutonept3}) removes the latter, the advanced propagator remains and is responsible for the dependence at advanced times seen in  (\ref{inoutquad1}).

To summarize, the in-out one-point function, as {\it derived from} $W$ in (\ref{inout0}) or (\ref{inout1}), is actually a matrix element of the field operator between the (in-)vacuum and a state $|Q\rangle$ that depends on the details of the evolution of the interacting field (see the definition of $|Q\rangle$ below (\ref{inoutonept2})). This matrix element is neither real nor causal and the final state of the system ($|0, {\rm out} \rangle = |0 \rangle$ here) is specified {\it ab initio} instead of being determined dynamically from a given initial state, as would be appropriate when given only initial data. 
 We will see below that a description of true expectation values is intimately related to an initial value formalism.

\subsection{In-in formalism for expectation values and causality}

To ensure that retarded boundary conditions are employed it is important to develop the appropriate generating functional. This is the so-called ``in-in" generating functional, which was first introduced by Schwinger \cite{Schwinger:JMathPhys2} and further developed by many others since (see 
Section \ref{sec:intro} for references). Here we review the construction of the ``in-in" effective action.

Let the in-in one-point function equal the true expectation value of the field so that
\begin{eqnarray}
	\langle \hat{\phi}_H (t, \bx) \rangle_{{\rm in-in}} 
		&=& \langle 0 | \hat{\phi}_H (t, \bx ) | 0 \rangle  \\
		&=& \langle 0 | \hat{U}_{Q} (-\infty, t) \hat{\phi}_I (t, \bx) \hat{U}_{Q} (t, -\infty) | 0 \rangle 
\end{eqnarray}
The first line implies that the one-point function on the left hand side is equivalent to the true expectation value of the field, 
and the last line shows that it is manifestly causal and real (Hermitian).
The subscript ``in-in" is included because the vacuum states that appear above are, more generally, the in-vacuum $| 0 , {\rm in} \rangle$, which gives it the name ``in-in." We seek to determine the functional whose derivative with respect to the auxiliary source yields these expressions. 

In analogy with (\ref{inout1}), a first guess might be
\begin{eqnarray}
	e^{ i W [ J ] } = \langle 0 | \hat{U}_{J+Q} (-\infty, \infty) \hat{U}_{J+Q} (\infty, -\infty) | 0 \rangle .  \label{inin0}
\end{eqnarray}
However, (\ref{inin0}) trivially equals one since the evolution operators are inverses of each other. Instead, we introduce an auxiliary source $J_1$ that couples to the field during the forward evolution in time and a second source $J_2$ for the evolution backward in time,
\begin{eqnarray}
	e^{ i W [ J_1, J_2 ] } = \langle 0 | \hat{U}_{J_2+Q} (-\infty, \infty) \hat{U}_{J_1+Q} (\infty, -\infty) | 0 \rangle  .\label{inin1}
\end{eqnarray}
Using (\ref{timeevo1}), the one-point functions are
\begin{eqnarray}
	\langle \hat{\phi}^1_H (t, \bx) \rangle _{{\rm in-in}} &\equiv& \frac{ \delta W}{ \delta J_1 (t, \bx) } \Bigg|_{J_1=J_2=0} = \langle 0 | \hat{U}_{Q} (-\infty, \infty) \hat{U}_{Q} (\infty, t) \hat{\phi}_I (t, \bx) \hat{U}_{Q} (t, -\infty) | 0 \rangle \label{ininonept0}  \\
	\langle \hat{\phi}^2_H (t, \bx) \rangle _{{\rm in-in}} &\equiv& - \frac{ \delta W}{ \delta J_2 (t, \bx) } \Bigg|_{J_1=J_2=0}  = \langle 0 |\hat{U}_{Q} (-\infty, t) \hat{\phi}_I (t, \bx) \hat{U}_{Q} (t, \infty)  \hat{U}_{Q} (\infty, -\infty)  | 0 \rangle     \label{ininonept1}
\end{eqnarray}
and equals the expectation value of the field in the vacuum state
\begin{eqnarray}
	\langle \hat{\phi}^1_H (t, \bx) \rangle _{{\rm in-in}} = \langle \hat{\phi}^2_H (t, \bx) \rangle _{{\rm in-in}}  = \langle 0 | \hat{\phi}_H (t, \bx) | 0 \rangle ,  \label{ininonept2}
\end{eqnarray}
as required.

The functional in (\ref{inin1}) has a path integral representation given by
\begin{eqnarray}
	e^{ i W [ J_1, J_2] } = \int {\cal D} \phi_1 {\cal D} \phi_2 \, \exp \Bigg\{ i S [ \phi_1 ] - i S [\phi_2 ] + i \int d^4x \, J_1 \phi_1 - i \int d^4x \, J_2 \phi_2 \Bigg\}  \label{inin4}
\end{eqnarray}
where $\phi_1 = \phi_2$ at $t =\infty$ according to (\ref{inin1}). This condition on the path integral contour provides an alternative name, the ``closed-time-path" or CTP formalism. We will use ``CTP" and ``in-in" interchangeably throughout the remainder. Notice that the presence of two sources implies path integrals over two fields, one each for the forward and backward time evolution depicted in (\ref{inin1}). 

For a linear scalar field, the path integrals can be evaluated exactly
\begin{eqnarray}
	W [ J_1, J_2 ] = \frac{ i}{2} \int d^4x \int d^4 y \, [ J_A (x) + Q_A (x) ]  G^{AB} (x-y) [ J_B (y) + Q_B(y) ]  ,   \label{inin6}
\end{eqnarray}
which can also be computed directly from (\ref{inin1}). Here $A,B = \{ 1,2 \}$, $Q_1 \equiv Q_2$ and
\begin{eqnarray}
	G^{AB} = \left( \begin{array}{cc}
				G_F & - G_- \\
				-G_+ & G_D
			\end{array} \right)  
\end{eqnarray}
where the quantities $G_{F,\pm,D}$ are given in Appendix \ref{app:twoptfns}. If $Q$ itself depends on a dynamical quantity (as in the next section where the quadrupole moment depends on the particle coordinates) then $Q_1 \ne Q_2$.

To check that this generating functional gives the appropriate causal structure for the one-point function, we compute the variation of $W$ with respect to one of the sources, using (\ref{ininonept0}) or (\ref{ininonept1}), and after using again the identities in Appendix \ref{app:twoptfns} find
\begin{eqnarray}
	\langle \hat{\phi}_H (t, \bx) \rangle _{{\rm in-in}} = \int d^4 y \, G_{ret} (x-y) Q(y) .   \label{onept0}
\end{eqnarray}
This demonstrates that the in-in one-point function manifestly satisfies retarded boundary conditions, which should be expected since (\ref{inin1}) depends on the initial state $|0, {\rm in} \rangle = | 0 \rangle$ only and not a final state {\it ab initio}. Indeed, the propagators $G^{AB}$ always combine to ensure that quantities of physical interest, such as equations of motion and expectation values, are manifestly {\it real} and {\it causal} \cite{Schwinger:JMathPhys2, Keldysh:JEPT20, Jordan:PRD33, CalzettaHu:PRD35}.

\subsection{The Keldysh representation}
\label{sec:keldysh}

The two-point functions appearing in the matrix $G^{AB}$ are not all independent, implying that we can isolate the fundamental ones by a suitable change of basis \footnote{The action $W$ is, of course, basis independent.}. This is accomplished by writing the sources in terms of their average $J_+$ and difference $J_-$,
\begin{eqnarray}
	J_- &=& J_1 - J_2 \\
	J_+ &=& \frac{ 1}{2} \big( J_1 + J_2 \big)  ,
\end{eqnarray}
so that
\begin{eqnarray}
	W [ J_+, J_- ] = \frac{ i}{2} \int d^4x \int d^4 y \, [ J_a (x) + Q_a (x) ]  G^{ab} (x-y) [ J_b (y) + Q_b (y) ]  .   \label{inin7}
\end{eqnarray}
Here $a,b = \pm$ and the matrix of two-point functions is given in the so-called Keldysh representation \cite{Keldysh:JEPT20} by
\begin{eqnarray}
	G^{ab} = \left( \begin{array}{cc}
				0 & - i G_{adv} \\
				-i G_{ret} & \frac{1}{2} G_H
			\end{array} \right)    \label{propmatrix0}
\end{eqnarray}
where $G^{++} = 0$ and $G_H$ is the symmetric (Hadamard) two-point function (see Appendix \ref{app:twoptfns}). The $a,b$ indices are raised and lowered with the ``metric"
\begin{eqnarray}
	c_{ab} = \left( \begin{array}{cc}
				0 & 1 \\
				1 & 0 
			\end{array} \right)  = c^{ab}   \label{ctpmetric0}
\end{eqnarray}
implying that $J_\pm = J^\mp$. The expectation value of the scalar field in the Keldysh basis is 
\begin{eqnarray}
	\langle \hat{\phi}_H (t, \bx) \rangle _{{\rm in-in}} &=& \langle \hat{\phi}_{H+} (t, \bx) \rangle _{{\rm in-in}} \big|_{J_\pm = 0}  = \frac{ \delta W}{ \delta J^+ } \Bigg|_{J_\pm = 0} = \frac{ \delta W}{ \delta J_- } \Bigg|_{J_\pm = 0} 
	\label{onept1}
\end{eqnarray}
and is easily shown to equal (\ref{onept0}) using (\ref{inin7}).

\section{The in-in formulation of NRGR}
\label{sec:ctp}

The retarded propagation of gravitational waves and their backreaction on the compact binary in the form of radiation reaction can be {\it self-consistently} and {\it systematically} implemented using the initial value formulation of quantum field theory in terms of the in-in effective action introduced in the previous section. We show how this effective action gives rise to the (classical) equations of motion for the binary system and how the emitted gravitational waves (i.e., the gravitational waveform) can be computed using the in-in framework within NRGR. We begin by perturbatively expanding the effective action and writing down the corresponding Feynman rules that allows one to circumvent the explicit evaluation of the path integrals.

Following (\ref{inin4}) a path integral representation for the in-in generating functional in NRGR is given by
\begin{eqnarray}
    e^{ i W [ \{ \bj_{K 1} \} , \{ \bj_{K 2} \} , J^{\mu\nu} _1, J^{\mu\nu}_2 ] } &=& \int \prod_{K=1}^2 {\cal D} \bx_{K 1} {\cal D} \bx_{K 2} {\cal D} \barh^{\mu\nu}_1 {\cal D} \barh^{\mu\nu}_2  \, \exp \Bigg\{ i S_{pot} [ \{ \bx_{K1} \} ,  \barh_1 ] -  iS_{pot} [ \{ \bx_{K2} \} , \barh_2 ] \nonumber \\
    && + i \sum_{K=1}^2 \int dt \big( {\bf j}_{K1} \cdot \bx_{K1} - {\bf j}_{K2} \cdot \bx_{K2} \big) +  i  \int d^4x \big( J_{1\mu\nu} \barh_1^{\mu\nu} - J_{2\mu\nu} \barh_2^{\mu\nu} \big) \Bigg\}    \label{genfun2}
\end{eqnarray}
where the conditions $\bx_{K1} = \bx_{K2}$ and $h_1^{\mu\nu} = h_2^{\mu\nu}$ are met at $t = \infty$. 
(Gauge-fixing in the in-in formalism follows the well-known Faddeev-Popov \cite{FaddeevPopov:PhysLett25} procedure and is applied to both graviton fields, $\barh^{\mu\nu}_1$ and $\barh^{\mu\nu}_2$.)

Perturbation theory in the in-in and in-out frameworks are similarly formulated. In particular, integrating out the radiation gravitons allows for (\ref{genfun2}) to be written as
\begin{eqnarray}
   e^{ i W [ \{ \bj_{K 1} \} , \{ \bj_{K 2} \} , J^{\mu\nu} _1, J^{\mu\nu}_2 ] } &\arreq&  \int \prod_{K=1}^2 {\cal D}\bx_{K1} {\cal D} \bx _{K2} \, \exp \Bigg\{ i \sum_{K=1}^2 \big( S_{pp} ^{(0)} [ \bx_{K1} ] -  S_{pp} ^{(0)} [ \bx_{K2}  ]  \big) + i \sum_{K=1}^2 \int dt \big( \bj_{K1} \cdot \bx_{K1} - \bj_{K2} \cdot \bx_{K2} \big) \nonumber \\
    && + i \int d^4x \, {\cal L} _{int} \Bigg[ \{ \bx_{K1} \}, \{ \bx_{K2} \}, -i \frac{ \delta }{ \delta J_1 ^{\alpha \beta} } , -i \frac{ \delta }{ \delta J_2 ^{\alpha \beta}} \Bigg] \Bigg\} Z_0 [J_1 ^{\mu\nu}, J_2^{\mu\nu} ]  ,    \label{genfun4}
\end{eqnarray}
which is expressed as a certain functional derivative operator acting on a Gaussian functional of the sources $J^{\mu\nu}_{1,2}$ and where the interaction Lagrangian is
\begin{eqnarray}
    \int d^4x \, {\cal L}_{int} &=& \sum_{n=1}^\infty \left( S_{pot}^{(n)} [ \{ \bx_{K1} \} , \barh_1 ] - S_{pot} ^{(n)} [ \{ \bx_{K2} \} , \barh_2 ]  \right) .
    \label{interactions1}
\end{eqnarray}
Here a superscript in parentheses denotes the number of radiation graviton fields contained in the interaction term. The quantity $Z_0$ is the free field generating functional for the radiation gravitons
\begin{eqnarray}
    Z_0 [ J_1 ^{\mu\nu}, J_2 ^{\mu\nu} ] &\!\!=\!\!& \! \int {\cal D} \barh_1^{\mu\nu} {\cal D} \barh_2 ^{\mu\nu} \, \exp \Bigg\{ i S^{(2)} [\barh_1] - i S^{(2)} [ \barh_2] + i \! \int \! d^4x \big( J_{1\mu\nu} \barh^{\mu\nu}_1 - J_{2 \mu\nu} \barh^{\mu\nu}_2 \big) \Bigg\} 
\end{eqnarray}
and is calculated by integrating the Gaussian along the CTP contour, which gives
\begin{eqnarray}
    Z_0 [ J_\pm ^{\mu\nu} ] = \exp \Bigg\{ - \frac{1}{2} \int d^4x  \int d^4 x^\prime \, J_a ^{\alpha \beta} (x) D_{\alpha \beta \gamma \delta} ^{ab} (x - x^\prime) J_b^{\gamma \delta} (x^\prime) \Bigg\}  .
\end{eqnarray}
In the Keldysh representation, the matrix of free graviton two-point functions is
\begin{eqnarray}
    D^{ab} _{\alpha \beta \gamma \delta} (x - x^\prime) = \left( \begin{array}{cc}
        0 & - i D^{adv} _{\alpha \beta \gamma \delta} \\
        - i D^{ret} _{\alpha \beta \gamma \delta} & \frac{1}{2} D^H _{\alpha \beta \gamma \delta}
        \end{array} \right)  \label{propmatrix1}
\end{eqnarray}
where $a,b = \pm$ and $D^{++}_{\alpha \beta \gamma \delta} = 0$. By construction, $W$ yields true expectation values of both $\hat{\bx}_K$ and $\hat{\barh}_{\mu\nu}$. 

Computing the (partial) Legendre transform of $W$ gives the in-in effective action
\begin{eqnarray}
    \Gamma [ \{ \langle \hat{\bx}_{K\pm} \rangle \} , J_\pm^{\mu\nu} ] &=& W [ \{ \bj_{K \pm} \} , J^{\mu\nu}_\pm ] - \sum_{K=1}^2 \int dt \, \bj_K^a \cdot \langle \hat{\bx}_{Ka} \rangle \\
    	&\equiv& S_{eff} [ \{ \langle \hat{\bx}_{K\pm} \rangle \} , J_\pm ^{\mu\nu} ]  .  \label{effaction0}   
\end{eqnarray}
The equations of motion for the true expectation values of the particle coordinates are then found by varying the effective action,
\begin{eqnarray}
    && 0 = \frac{ \delta \Gamma }{ \delta \langle \hat{\bx}_{K-} \rangle } \Bigg| _{  \bx_{K-} = 0, \, \bx_{K+} = \bx_K, \, \bj_{K\pm} = J_\pm^{\mu\nu} = 0   }  .\label{particleeom0} 
\end{eqnarray}
One can show that the large masses of the compact objects and the influence from graviton quantum fluctuations sufficiently decohere the coordinate operators $\hat{\bx}_K$ and ensure that the above equations of motion for the expectation value appropriately describe the classical limit of the compact binary. See \cite{Johnson:PhD, JohnsonHu:PRD65, GalleyHu:PRD72, GalleyHuLin:PRD74} and references therein for elaboration of this point.
The gravitational waves radiated by the binary can be computed from the graviton in-in one-point function (see (\ref{onept1}))
\begin{eqnarray}
	\langle \hat{\barh}_{\mu\nu} (t, \bx) \rangle = \frac{\delta W}{ \delta J_-^{\mu\nu} (t, \bx) } \Bigg| _{  \bx_{K-} = 0, \, \bx_{K+} = \bx_K, \, \bj_{K\pm} = J_\pm^{\mu\nu} = 0   }     \label{ininonept00}
\end{eqnarray}  
and forms the vacuum expectation value of the radiation graviton field.

One can generally show that (\ref{particleeom0}) yields equations of motion for the trajectories of the compact objects that are manifestly {\it real} and {\it causal}. Furthermore, (\ref{ininonept00}) is a solution to the field equations for the radiation modes and is guaranteed to satisfy retarded boundary conditions. See \cite{CalzettaHu:PRD35, CalzettaHu:PRD37} for proofs of these statements. See also \cite{CalzettaHu} for a more thorough discussion of the in-in/CTP formalism and its use for describing dissipative and, more generally, nonequilibrium processes in quantum field theory and quantum mechanics.

\subsection{NRGR Feynman rules in the in-in formalism}
\label{sec:feynrules}

\begin{figure}
	\center
	\includegraphics[width=3.75cm]{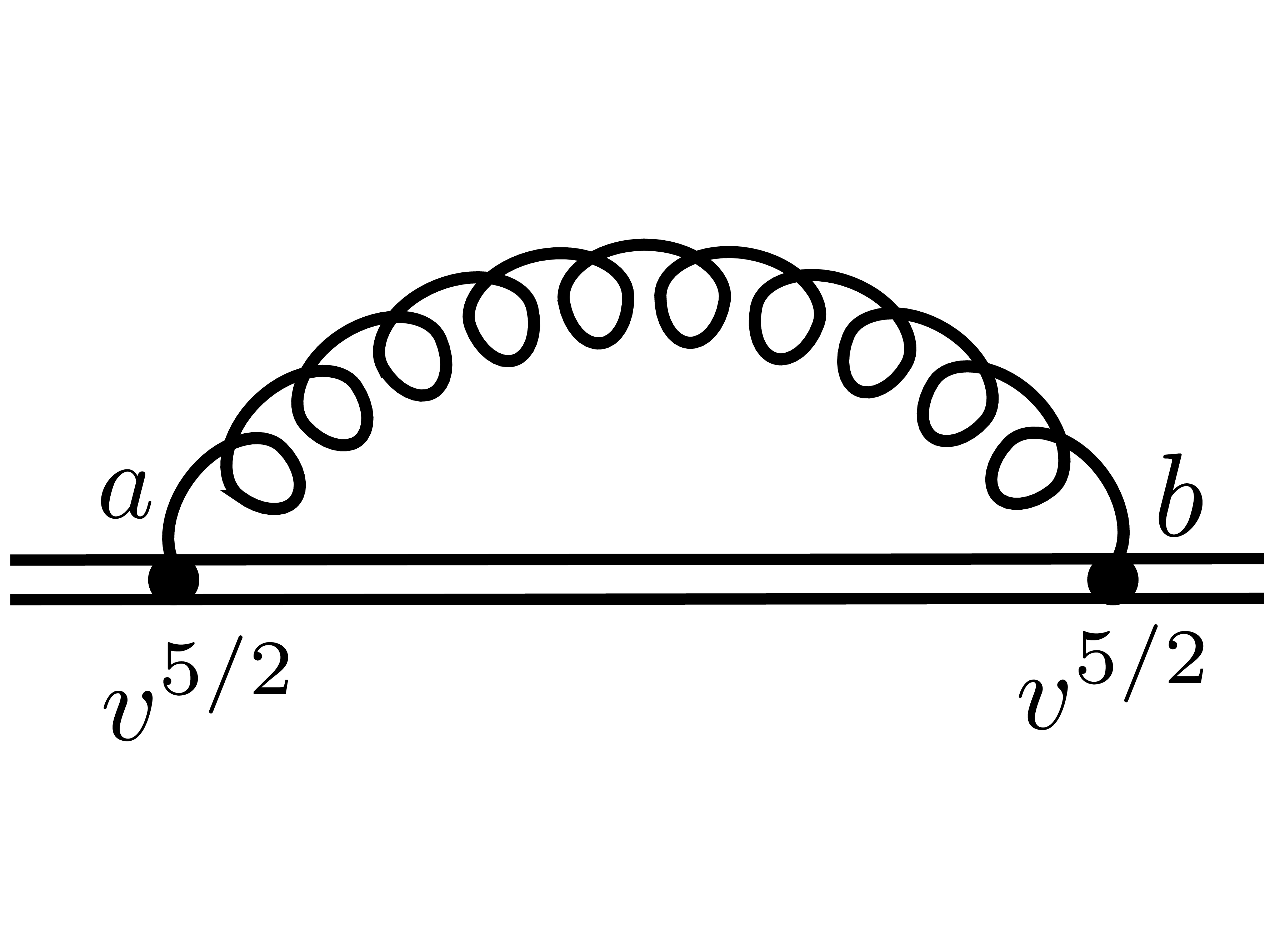}
	\caption{The non-zero diagram contributing to the leading order (2.5PN) radiation reaction in NRGR using the in-in formalism. Note the CTP indices $a,b$ that relate to the forward and backward branches of time in the closed-time-path contour of the CTP path integral.}
	\label{fig:ininLOradrxn}
\end{figure}

The diagrammatic structure of in-in perturbation theory is nearly identical to the in-out approach. However, when drawing Feynman diagrams in the in-in formalism it is necessary to include CTP labels $a,b,c,\ldots$ at each vertex to keep track of the forward and backward branches of time in the CTP path integral. For example, the leading order contribution to radiation reaction in the in-in approach is given in Fig.(\ref{fig:ininLOradrxn}). 

The Feynman rules for calculating the in-in effective action (e.g., for deriving radiation reaction forces) are similar to those given in \cite{GoldbergerRothstein:PRD73}:
\begin{itemize}
	\item Include a factor of the radiation graviton two-point function $D^{ab}_{\alpha \beta \gamma^\prime \delta^\prime} (x,x^\prime)$ connecting particle-field vertices labeled by CTP indices $a$ and $b$ at spacetime points $x$ and $x^\prime$,
	\item A vertex coupled to $n$ radiation gravitons is labelled with $n$ CTP indices, one for each graviton,
	\item Sum over all CTP indices.
\end{itemize}
\noindent When computing the one-point function of the radiation graviton there is an additional rule for those graviton lines that connect to a distant field point. In the Keldysh representation:
\begin{itemize}
	
	\item Include a factor of $D^{-a}_{\alpha \beta \gamma^\prime \delta^\prime} (x,x^\prime)$ for each radiation graviton that connects a vertex at event $x^\prime$ with CTP index $a$ to the {\it field} point $x$.
	
\end{itemize}
In the following sections, we use these rules to derive the expressions for the leading order radiation reaction and gravitational wave emission. Initially, we use the matrix of propagators given in (\ref{propmatrix1}) to demonstrate these rules. However, one can equivalently use (\ref{propmatrix12}) instead for all tree-level diagrams since (\ref{propmatrix12}) contains only the relevant propagators for constructing the effective action and graviton $1$-point functions in the classical limit as we show below.

\subsection{Radiation reaction}

The diagram for the leading order radiation reaction is given in Fig.(\ref{fig:ininLOradrxn}). From the Feynman rules above we write down the term in the 2.5PN effective action that corresponds to Fig.(\ref{fig:ininLOradrxn}),
\begin{eqnarray}
	i S_{eff}^{2.5 {\rm PN} } [ \{ \bx_{K\pm} \} ] = \left( \frac{1}{2} \right) \left( \frac{i}{ 2m_{pl} } \right)^2 \int dt \int dt^\prime \, Q^{ij}_a (t) \big\langle R_{0i0j}^a (t) R_{0k0l}^b (t^\prime) \big\rangle Q^{kl} _b (t^\prime)   \label{ininrr1}
\end{eqnarray}
where we are using the summation convention for the CTP indices $a,b=\pm$.
The two-point function of the product of linearized Riemann tensors is as in (\ref{prodRiem1}) but replacing the Feynman propagator there by the in-in two-point functions (\ref{propmatrix1}) so that
\begin{eqnarray}
	\big\langle R_{0i0j}^a (t) R_{0k0l}^b (t^\prime) \big\rangle = \frac{1}{20} \Bigg[ \delta_{ik} \delta_{jl} + \delta_{ik} \delta_{jl} - \frac{2}{3} \delta_{ij} \delta _{kl} \Bigg] \frac{d^2 }{ dt^2 } \frac{ d^2 }{ dt^{\prime 2} } D^{ab} (t-t^\prime, \bzero)
\end{eqnarray}
where $D^{ab}_{ijkl} = P_{ijkl} D^{ab }$ defines $D^{ab}$ and
\begin{eqnarray}
	P_{\alpha \beta \gamma \delta} = \frac{1}{2} \big( \eta_{\alpha \gamma} \eta_{\beta \delta} + \eta_{\alpha \delta} \eta_{\beta \gamma} - \eta_{\alpha \beta} \eta_{\gamma \delta} \big)
\end{eqnarray}
in four spacetime dimensions.
Performing the CTP summations gives
\begin{eqnarray}
	i S_{eff} ^{2.5 {\rm PN} } [ \{ \bx_{K\pm} \} ] &\arreq& - \frac{1}{80 m_{pl}^2 } \int dt \int dt^\prime \left[ - 2 i Q^{ij}_- (t) \frac{ d^2 }{ dt^2 } \frac{ d^2 }{ dt^{\prime 2} } D_{ret} (t- t^\prime, \bzero) Q^{ij}_+ (t^\prime) + \frac{1}{2} Q_- ^{ij} (t) \frac{d^2 }{ dt^2} \frac{ d^2 }{ dt^{\prime 2}} D_H (t-t^\prime, \bzero) Q^{ij}_- (t^\prime) \right]  \nonumber \\
	&& \label{ininrr2}
\end{eqnarray} 
where the differenced and averaged quadrupole moments are
\begin{eqnarray}
	Q_- ^{ij} (t) &\equiv& Q^{ij}_1 (t  ) - Q^{ij}_2 (t ) \\
	Q_+^{ij} (t) &\equiv& \frac{1}{2} \Big( Q^{ij}_1 (t ) + Q^{ij}_2 (t ) \Big)  
\end{eqnarray}
and $Q_{1,2}^{ij}$ are the quadrupole moments for the worldline histories $\{ \bx_K^{1,2} \}$.
In terms of the $\bx_{K\pm}$ variables one can show that 
\begin{eqnarray}
	Q_- ^{ij} (t) &=& \sum_K m_K \left\{ x_{K-}^i x_{K+}^j + x_{K+}^i x_{K-}^j - \frac{2}{3} \delta^{ij} \bx_{K-} \cdot \bx_{K+} \right\}  \\
	Q_+^{ij} (t) &=& \sum_K m_K \left\{ x_{K+}^i x_{K+}^j - \frac{1}{3} \delta^{ij} \bx_{K+}^2 \right\} + O( x_-^3 )  .
\end{eqnarray}
According to (\ref{particleeom0}) we can ignore the last term in (\ref{ininrr2}) and the $O(x_-^3)$ terms from $Q_+^{ij}$ since these do not contribute to the equations of motion.
In fact, the $O(\bx_-^2)$ terms are related to the effects that the quantum fluctuations of the radiation graviton field have on the 
trajectories. This is discussed in more detail for the case of a particle moving in a curved background spacetime in \cite{GalleyHu:PRD72, GalleyHuLin:PRD74}. We will ignore these higher order terms throughout the remainder since they do not contribute to the equations of motion or the gravitational wave emission in the classical limit. In fact, in the Feynman rules we can (and from here on we do) simply use the following matrix of propagators
\begin{eqnarray}
	D^{ab}_{\alpha \beta \gamma \delta} (x-x^\prime) = \left( \begin{array}{cc}
												0 & - i D^{adv}_{\alpha \beta \gamma \delta} \\
												- i D^{ret}_{\alpha \beta \gamma \delta} & 0 
											\end{array} \right) 
	\label{propmatrix12}
\end{eqnarray}
instead of (\ref{propmatrix1}).

The 2.5PN effective action can then be written as
\begin{eqnarray}
	i S_{eff}^{2.5 {\rm PN} } [ \{ \bx_{K\pm} \} ] &=& \frac{ i }{20 m_{pl}^2 } \int dt \int dt^\prime \,  \frac{d^2 }{ dt^2 } \frac{ d^2 }{ dt^{\prime 2} } D_{ret} (t-t^\prime, \bzero) \sum_{K=1}^2 m_K x_{K-}^i (t) x_{K+}^j (t) Q^{ij}_+ (t^\prime)   .   \label{ininrr23}
\end{eqnarray}
where we ignore the $O(x_-^3)$ terms in $Q_+^{ij}$.
Using
\begin{eqnarray}
	D_{ret} (t-t ^\prime , \bzero) = - \frac{ 1}{2 \pi} \theta ( t- t^\prime ) \delta \big( (t- t^\prime)^2 \big)
\end{eqnarray}
and dimensional regularization one can show that
\begin{eqnarray}
	\int dt^\prime \, \frac{d^2 }{ dt^2 } \frac{ d^2 }{ dt^{\prime 2} } D_{ret} (t-t^\prime, \bzero) Q^{ij}_+ (t^\prime) = - \frac{ 1}{4\pi} \frac{d ^5 Q^{ij}_+ (t) }{ dt^5}   ,
\end{eqnarray}
from which the effective action follows
\begin{eqnarray}
	i S_{eff}^{2.5 {\rm PN} } [ \{ \bx_{K\pm} \} ] &=& - \frac{ i }{80 \pi m_{pl}^2 } \int dt \sum_{K=1}^2 m_K x_{K-}^i (t) x_{K+}^j (t) \frac{d ^5 Q^{ij}_+ (t) }{ dt^5}   \label{ininrr3}
\end{eqnarray}
and provides the following term to the equations of motion
\begin{eqnarray}
	\frac{ \delta S_{eff} ^{2.5 {\rm PN} } }{ \delta x^i_{K-} (t) } \Bigg|_{\bx_{K-} = 0, \, \bx_{K+} = \bx_K } {\hskip-0.2in} = - \frac{ 2 m_K G_N }{ 5 } \, \frac{d ^5 Q_{ij} (t) }{ dt^5} x_K^j (t)  \label{burkethorne1}
\end{eqnarray}
with $m_{pl}^{-2} = 32 \pi G_N$. We observe that this is precisely the radiation reaction force on the $K^{\rm th}$ body obtained by Burke and Thorne \cite{BurkeThorne:Relativity, Thorne:AstrophysJ158, Burke:JMathPhys12, Maggiore}.

\subsection{Quadrupole radiation}

The diagram for the leading order (quadrupole) radiation is given in (\ref{radiation0}) with the addition of the CTP labels ``$-$" and ``$a$" at the free end of the radiation graviton and at the vertex, respectively, so that
\begin{eqnarray}
	\barh_{ij} ^{TT} (t, \bx) &=& \left( \frac{ i }{ 2m_{pl}} \right) \Lambda _{ij,kl} (\hat{\bn} ) \int dt^\prime \, \left[ \frac{1}{2} D_{klmn}^{-a} (t-t^\prime, \bx) \right] \ddot{Q}_a^{mn} (t^\prime) \Bigg|_{\bx_{K-} = 0, \, \bx_{K+} = \bx_K }
\end{eqnarray}
in the TT gauge. The factor of $1/2$ in the square brackets comes from the linearized Riemann tensor.
Only the $a=+$ term contributes in the CTP summation so that
\begin{eqnarray}
	m_{pl}^{-1} \barh_{ij} ^{TT} (t, \bx) = \frac{ 1 }{ 4 m_{pl}^2 } \Lambda _{ij,kl} \int dt^\prime \, D^{ret}_{klmn} (t-t^\prime, \bx) \ddot{Q}^{mn} (t^\prime),
\end{eqnarray}
with
\begin{eqnarray}
	D^{ret}_{klmn} (t-t^\prime, \bx) = - \theta ( t-t^\prime) \frac{ \delta ( t-t^\prime - | \bx | ) }{ 4 \pi | \bx | } P_{klmn} .
\end{eqnarray}
It follows that the quadrupole radiation is causally propagated from the source,
\begin{eqnarray}
	m_{pl}^{-1} \barh_{ij} ^{TT} (t, \bx) = - \frac{ 2 G_N}{ | \bx | } \Lambda _{ij,kl} \ddot{Q}_{kl} (t - | \bx | )   .\label{iningw2}
\end{eqnarray}
Notice that (\ref{iningw2}) is purely real and causal, which is guaranteed by the in-in formalism, and that the causal and time-asymmetric emission of gravitational waves is consistent with the presence of radiation reaction forces in (\ref{burkethorne1}).

From (\ref{iningw2}) one can compute the leading order contribution in a multipole expansion to the power radiated per unit solid angle
\begin{eqnarray}
	\frac{dP}{d \Omega} = |\bx|^2 \big\langle \dot{\barh}_{ij}^{TT} (t, \bx) \dot{\barh}_{ij}^{TT} (t, \bx) \big\rangle 
\end{eqnarray}
from which the total power is given by
\begin{eqnarray}
	P = \frac{G_N}{5} \big\langle \dddot{Q}_{ij} (t - |\bx| ) \dddot{Q}_{ij} (t - | \bx|) \big\rangle + \cdots
\end{eqnarray}
where here the angled brackets denote a temporal average over several periods of the gravitational wave and $\cdots$ represents higher order contributions in the multipole expansion. Notice that the power depends on the quadrupole moment of the binary at retarded times. 

Interestingly, the multipole expansion of the power loss and change in time of the linear and angular momenta can also be computed directly in the in-in approach using Feynman diagrams. Indeed, the expectation value of the graviton stress tensor is related to the coincidence limit of the Hadamard two-point function, $D^H_{\alpha \beta \gamma \delta} (x,x')$ \cite{BirrellDavies}, and is in turn related to the imaginary part of the effective action in (\ref{ininrr2}).

\section{Conclusions}

In this paper we computed the leading order radiation reaction forces on the compact objects within NRGR using an initial value formulation of the underlying path integral framework to implement retarded boundary conditions on the radiated gravitational perturbations. We also calculated the quadrupole radiation emitted causally by the compact binary and the leading order contribution to the radiated power. We showed that using the in-out framework is not suitable for describing the retarded propagation of true expectation values of radiation gravitons and their backreaction on the compact objects in the form of radiation reaction. 
Using the in-in formalism guarantees real and causal $n$-point functions and equations of motion for expectation values. 

Whereas the in-out formalism is usefully applicable for the practical purposes of describing 
the conservative forces appearing in the equations of motion for the compact objects and the instantaneous power emitted in gravitational waves, the in-in formalism is necessary for self-consistently deriving the radiation reaction forces on the compact objects and the multipole expansion of gravitational waves propagating causally in the far-zone, among other things. The self-consistent nature of the in-in approach to NRGR should also be useful for computing the hereditary contributions to the power loss and radiation field.

In Appendix \ref{sec:cleft} we discussed the equivalence of NRGR and the classical effective field theory (ClEFT) approach of Kol and Smolkin \cite{KolSmolkin:PRD77}. We also showed that using only retarded propagators does not lead to a consistent theory in the original formulation of \cite{KolSmolkin:PRD77}. 
 We provided a new formulation of ClEFT, which is simply an alternative derivation of the in-in approach to NRGR, that is appropriate for the radiation sector and consistently implements retarded boundary conditions within a strictly classical framework. This is achieved by doubling the degrees of freedom in a manner that is equivalent to the in-in formulation of NRGR.
The appearance of advanced propagators at intermediate steps is necessary to ensure the self-consistency and causality of the derived equations of motion and gravitational waves.

\acknowledgments
CRG sincerely thanks Ted Jacobson and Ira Rothstein for a critical reading of a previous draft. He also thanks Ira Rothstein for many engaging and enlightening discussions on the applicability of in-out and in-in techniques. We also thank Barak Kol for discussions. This research has been supported by NSF Grant No. 0801213 to the University of Maryland.

\appendix

\section{Coupling to conserved quantities}
\label{app:zerodiagram}

\begin{figure}
	\center
	\includegraphics[width=4cm]{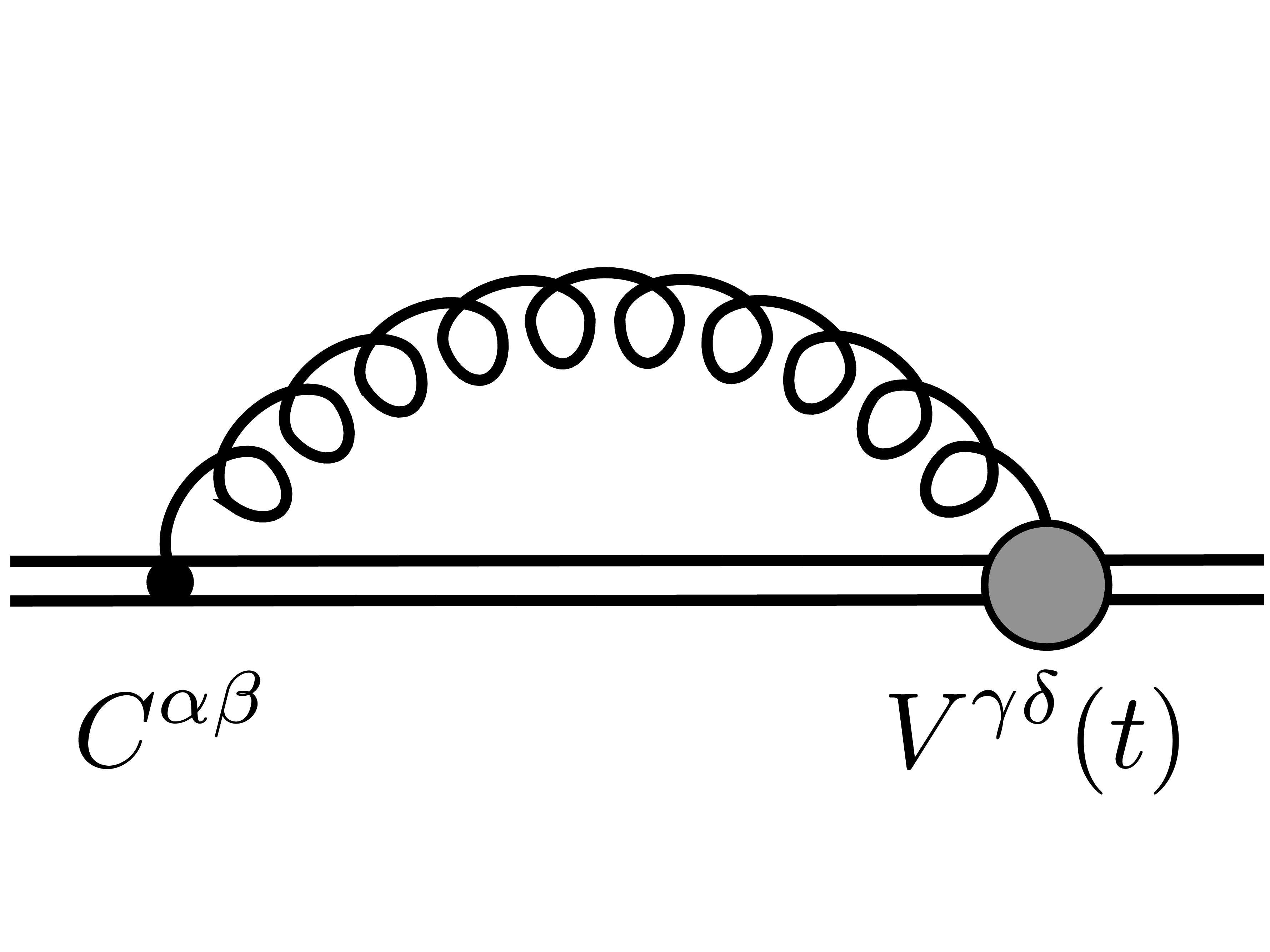}
	\caption{The coupling of a radiation graviton to a time-independent vertex $C^{\alpha \beta}$ and a general, time-dependent vertex $V^{\alpha \beta}(t)$, represented here as a gray circle.}
	\label{fig:zerodiagram}
\end{figure}

In this Appendix we show that the coupling of a generic vertex $V^{\mu\nu} (t)$ to a conserved quantity $C^{\alpha \beta}$ gives a vanishing contribution to the equations of motion. Applying the (in-out) Feynman rules to the diagram in Fig.(\ref{fig:zerodiagram}) gives
\begin{eqnarray}
	{\rm Fig.(\ref{fig:zerodiagram})} = i S = \int _{-\infty}^\infty \!\!\! dt \int _{-\infty}^\infty \!\!\! dt^\prime \, i C^{\alpha \beta} D^F_{\alpha \beta \gamma \delta} (t-t^\prime, \bzero) \, i V^{\gamma \delta}  (t^\prime),
\end{eqnarray}
where $C^{\alpha \beta}$ is time-independent to the given order in velocity. The vertex $V^{\gamma \delta} (t^\prime)$ could be built out of other radiation gravitons but its exact structure is irrelevant for this general discussion. The momentum space representation for the Feynman propagator gives
\begin{eqnarray}
	i S = - C^{\alpha \beta} P_{\alpha \beta \gamma \delta} \int dt  \int dt^\prime \int_{k^0, \bk} \frac{ i e^{ - i k^0 (t-t^\prime) } }{ (k^0)^2 - \bk^2 + i \epsilon} \,  V^{\gamma \delta} (t^\prime)   . \label{zero0}
\end{eqnarray}
Using the distributional identity \cite{CardosoDiasFigueras:PRD78}
\begin{eqnarray}
	\int d k^0 \, \frac{ f( k^0) }{ (k^0)^2 - \bk^2 + i \epsilon } = PV \int dk^0  \, \frac{ f( k^0) }{ (k^0)^2 - \bk^2 } + i \pi \int dk^0 \, f(k^0) \delta \big( (k^0)^2 - \bk^2 \big)  ,  \label{PV0}
\end{eqnarray}
where $PV$ denotes the principal value of the integral and $f(k^0)$ is an arbitrary function, yields an expression more amenable for isolating the real and imaginary parts of $S$
\begin{eqnarray}
	i S = - i C^{\alpha \beta} P_{\alpha \beta \gamma \delta} \int dt \int dt^\prime \int_{\bk} \left\{ PV \int_{k^0} \frac{ e^{-i k^0 (t-t^\prime)}  }{ (k^0)^2 - \bk^2} V^{\gamma \delta} (t^\prime) + i \pi \int_{k^0} \, e^{- i k^0 ( t- t^\prime) } \delta \big( (k^0)^2 - \bk^2 \big) V^{\gamma \delta}(t^\prime) \right\}   .
\end{eqnarray}
In the first term, define $s= t^\prime -t$ and expand $V^{\gamma \delta} (t+s)$ in powers of $s$ so that
\begin{eqnarray}
	i S &=& - i C^{\alpha \beta} P_{\alpha \beta \gamma \delta} \int dt \Bigg\{ \sum_{n=0}^\infty \frac{ 1}{n!} \frac{ d^n }{ dt^n} V^{\gamma \delta} (t) \int_{-\infty}^\infty \!\!\! ds \, \int _{\bk} PV \int_{k^0} \, \frac{ s^n  e^{i k^0 s} }{ (k^0)^2 - \bk^2 } + i \pi \int_{\bk} \int_{k^0} e^{-i k^0 t} \delta \big( (k^0)^2 - \bk^2 \big) \tilde{V}^{\gamma \delta} (k^0) \Bigg\} \nonumber \\
	&&  \label{zero1}
\end{eqnarray}
where the Fourier transform of $V^{\mu\nu} (t^\prime)$ is
\begin{eqnarray}
	\tilde{V}^{\mu\nu} (k^0) = \int_{-\infty}^\infty dt^\prime \, e^{i k^0 t^\prime} V^{\mu\nu} (t^\prime)  .  \label{fourier1}
\end{eqnarray}
From the discussion in Appendix \ref{app:master}, the only contribution from the master integral to the first term in (\ref{zero1}) occurs when $n=1$. Also, the second term vanishes entirely since
\begin{eqnarray}
	\int dt \, \int _{\bk} \int_{k^0} e^{-i k^0 t} \delta \big( (k^0)^2 - \bk^2 \big) \tilde{V}^{\gamma \delta } (k^0) = \tilde{V}^{\gamma \delta} ( 0 ) \int_{\bk} \delta \big( \bk ^2 \big) = 0   .
\end{eqnarray}
Therefore, the contribution to the effective action from a diagram with the structure given in Fig.(\ref{fig:zerodiagram}) is
\begin{eqnarray}
	S = - C^{\alpha \beta} P_{\alpha \beta \gamma \delta} \int dt \, \frac{ d V^{\gamma \delta} }{ dt}  ,
\end{eqnarray}
which is a total time derivative and does not contribute to the equations of motion, as claimed. It is a straightforward matter to extend the calculation here to the in-in case.

\section{Master integral}
\label{app:master}

The master integral we use in the main body of the text is defined as
\begin{eqnarray}
	I ( n, p, q) \equiv \int_{-\infty} ^\infty ds \int_{\bk} PV \int_{k^0}  \frac{ e^{i k^0 s} }{ (k^0)^2 - \bk^2 } \, s^n (k^0)^p k_{i_1} \cdots k_{i_q}  \label{master0}
\end{eqnarray}
and is non-zero when $q$ is an even integer. Writing $s^n$ as $n$ derivatives of $e^{ik^0s}$ with respect to $k^0$ then integrating over $s$ and $k^0$ gives
\begin{eqnarray}
	I (n, p, q) = i^n \left\{ \frac{ d^n}{ d(k^0)^n } \left[ (k^0)^p \int_{\bk} \frac{ k_{i_1} \cdots k_{i_q} }{ (k^0)^2 - \bk^2 } \right] \right\}_{k^0=0}  .
\end{eqnarray}
Under the $\bk$ integral the product of momentum vector components can be written as
\begin{eqnarray}
	k_{i_1} \cdots k_{i_q} &=& \frac{ | \bk | ^q }{ (q+1)!! } \big( \delta_{i_1 i_2} \cdots \delta_{i_{q-1} i_q} + \cdots \big)  \\
		&\equiv& \frac{ | \bk | ^q }{ (q+1)!! } \, \delta_{i_1 \cdots i_q}
\end{eqnarray}
where the quantity in parentheses represents all possible pairings of the $q$ indices \cite{Maggiore}. In spherical coordinates, the $\bk$ integral becomes
\begin{eqnarray}
	\int_{\bk} = \int _{-\infty}^\infty \frac{ d^{d-1} k}{ (2\pi)^{d-1} } = \frac{1}{ (2 \pi)^{d-1} } \int_{S^{d-2}} {\hskip-0.15in} d\Omega \int_0 ^\infty dk \, k^{d-2}
\end{eqnarray}
and integrating over the $d-2$ dimensional sphere gives
\begin{eqnarray}
	I (n, p, q) = \frac{ i^n \delta _{i_1 \cdots i_q} }{ (q+1)!! \, 2^{d-2} \pi ^{\frac{d-1}{2}} \Gamma \left( \frac{ d-1}{2} \right) } \left\{ \frac{ d^n }{ d(k^0)^n } \left[ (k^0)^p \int_0 ^\infty dk \, \frac{ k^{d-2+q} }{(k^0)^2 - \bk^2 } \right] \right\} _{k^0=0}  .
\end{eqnarray}
Using dimensional regularization to evaluate the remaining integral and computing the $k^0$ derivatives on the resulting expression yields
\begin{eqnarray}
	I ( n, p, q) = \frac{ i^{n-d+3} \sec \left( \frac{ \pi d}{2} \right) }{ 2^{d-1} \pi^{\frac{d-3}{2} } \Gamma \left( \frac{ d-1}{ 2} \right) } \frac{\delta_{i_1 \cdots i_q} }{ (q+1)!! } \frac{ ( p +q+d-3)! }{ ( p + q + d-3-n)!} (k^0)^{p+q+d-3-n} \Big|_{k^0=0}   .
\end{eqnarray}
Notice that there is a non-zero contribution only when $p+q+d-3=n$ implying that the master integral in $d$ spacetime dimensions is
\begin{eqnarray}
	I( p + q + d-3, p, q) = \frac{ i^{n-d+3} \sec \left( \frac{ \pi d}{2} \right) }{ 2^{d-1} \pi^{\frac{d-3}{2} } \Gamma \left( \frac{ d-1}{ 2} \right) } \frac{\delta_{i_1 \cdots i_q} }{ (q+1)!! } \, n!   
\end{eqnarray}
and in $d=4$ dimensions becomes
\begin{eqnarray}
	I ( p+q+1, p, q) = \frac{ i^{n-1} n!}{ 4 \pi (q+1)!! } \, \delta_{i_1 \cdots i_q}   .  \label{master1}
\end{eqnarray}
For $n \ne p+q+d-3$, the master integral vanishes.

\section{Propagators and two-point functions}
\label{app:twoptfns}

In this Appendix we collect some definitions, identities and relations for the quantum two-point functions that are relevant for this work. 

The positive and negative frequency Wightman functions are defined as
\begin{eqnarray}
    D_{\alpha \beta \gamma^\prime \delta^\prime} ^+ (x,x^\prime) &\arreq& \big\langle \hath_{\alpha \beta} (x) \hath _{\gamma^\prime \delta^\prime} (x^\prime) \big\rangle  \label{posfreq0} \\
    D_{\alpha \beta \gamma^\prime \delta^\prime} ^- (x,x^\prime) &\arreq& \big \langle \hath _{\gamma^\prime \delta ^\prime} (x^\prime) \hath _{\alpha \beta} (x) \big\rangle  ,  \label{negfreq0}
\end{eqnarray}
respectively. The Feynman, Dyson, Hadamard and commutator two-point functions are, respectively,
\begin{eqnarray}
    D_{\alpha \beta \gamma^\prime \delta^\prime} ^F (x,x^\prime) &=& \big\langle T \, \hath_{\alpha \beta}(x) \hath_{\gamma^\prime \delta^\prime} (x^\prime) \big\rangle  \label{feynman0} \\
    D_{\alpha \beta \gamma^\prime \delta^\prime} ^D (x,x^\prime) &=& \big\langle T^* \, \hath_{\alpha \beta}(x) \hath_{\gamma^\prime \delta^\prime} (x^\prime) \big\rangle  \label{dyson0} \\
    D_{\alpha \beta \gamma^\prime \delta^\prime} ^H (x,x^\prime) &=& \big\langle \{ \hath_{\alpha \beta} (x), \hath _{\gamma^\prime \delta^\prime} (x^\prime) \} \big\rangle \label{hadamardfn0} \\
    D_{\alpha \beta \gamma^\prime \delta^\prime} ^C (x,x^\prime) &=& \big\langle [ \hath_{\alpha \beta} (x), \hath_{\gamma^\prime \delta^\prime} (x^\prime) ] \big\rangle  \label{commutator0}
\end{eqnarray}
where $T$ is the time-ordering operator and $T^*$ is the anti-time-ordering operator. The field commutator is independent of the particular state used to evaluate it. Given the Wightman functions in (\ref{posfreq0}) and (\ref{negfreq0}) we write the above two-point functions in the form (ignoring the tensor indices from here on)
\begin{eqnarray}
    D_F (x,x^\prime) &\arreq& \theta( t-t^\prime) D_+ (x,x^\prime) + \theta( t^\prime-t) D_- (x,x^\prime) \label{feynman1}  \\
    D_D (x,x^\prime ) &\arreq& \theta( t^\prime-t) D_+ (x,x^\prime) + \theta( t-t^\prime) D_- (x,x^\prime) \label{dyson1}    \\
    D_H (x,x^\prime) &\arreq& D_+ (x,x^\prime) + D_- (x,x^\prime)  \label{hadamardfn1}  \\
    D_C (x,x^\prime) &\arreq& D_+ (x,x^\prime) - D_- (x,x^\prime)  . \label{commutator1}
\end{eqnarray}
From these we define the retarded and advanced propagators by
\begin{eqnarray}
    -i D_{ret} (x,x^\prime ) &\arreq&  \theta (t-t^\prime) D_C (x,x^\prime) \label{retardedfn0} \\
    +i D_{adv} (x,x^\prime) &\arreq&  \theta(t^\prime-t) D_C(x,x^\prime)  . \label{advancedfn0}
\end{eqnarray}
These propagators also satisfy the following useful identities
\begin{eqnarray}
    -i D_{ret} (x,x^\prime ) &=& D_F (x,x^\prime ) - D_- (x,x^\prime ) \\
    		&=& D_+ (x,x^\prime ) - D_D (x,x^\prime ) \\
    i D_{adv} (x,x^\prime ) &=& D_D (x,x^\prime ) - D_- (x,x^\prime ) \\
    		&=& D_+ (x,x^\prime ) - D_F (x,x^\prime )   
\end{eqnarray}
from which the Feynman propagator can be written in terms of its real and imaginary parts as
\begin{eqnarray}
	D_F (x,x^\prime) &=& - \frac{i}{2} \big[ D_{ret} (x,x^\prime) + D_{adv} (x,x^\prime) \big]  - \frac{1}{2} D_H(x,x^\prime)   .  \label{feynmanRI}
\end{eqnarray}
The Feynman, Dyson and Wightman functions are not all independent since
\begin{eqnarray}
    D_H (x,x^\prime ) &=& D_F (x,x^\prime ) + D_D (x,x^\prime ) \\
    		&=& D_+ (x,x^\prime ) + D_- (x,x^\prime )  .
\end{eqnarray}
Under the interchange of $x$ and $x^\prime$ the Feynman, Dyson and Hadamard two-point functions are symmetric, the commutator is anti-symmetric and
\begin{eqnarray}
    D_+ (x,x^\prime) &=& D_- (x^\prime, x) \\
    D_{ret} (x,x^\prime) &=& D_{adv} (x^\prime, x)  .
\end{eqnarray}

\section{Gravitational waves in classical effective field theory}
\label{sec:cleft}

The classical effective field theory (ClEFT) approach, introduced in \cite{KolSmolkin:PRD77} and developed further in \cite{KolSmolkin:CQG25, CardosoDiasFigueras:PRD78, Kol:GRG40, Levi:0802.1508}, is an attempt to simplify NRGR \cite{KolSmolkin:PRD77} by using only classical methods and techniques. More specifically, ClEFT uses Newton's constant $G_N$ in the power counting instead of the Planck mass $m_{pl}^{-2} = 32 \pi G_N$, vertices and propagators are defined without factors of $i$, integrating out gravitons from the theory does not use non-rigorous path integral methods, and radiation graviton lines in Feynman diagrams represent retarded propagators since one is working within a classical field theory with retarded boundary conditions \footnote{We thank Barak Kol for making this point clear to us in a private communication.}. The first two differences are superficial since these are related to subjective conventions 
\footnote{For example, to demonstrate that it is acceptable to power count the (classical) interactions of NRGR using the Planck mass choose units where only $c=1$ but $\hbar \ne 1$. Then the leading order particle-graviton interaction is $S_{pp} \sim m \int dt (\hbar^{1/2}H_{00} / m_{pl} )$ and the effective action in the classical limit at $O(v^0L)$ is $S_{eff}/ \hbar \sim  m^2 / m_{pl}^2 / v$. One can show that $m^2/m_{pl}^2 \sim v L /\hbar$ and so $S_{eff} \sim v^0 L$ and $\hbar$ does not appear.}.
The third attempts to eliminate the use of path integrals and to integrate out gravitons by solving for the short wavelength perturbations on the long wavelength classical background, then evaluating the resulting action \cite{KolSmolkin:PRD77}. Of course, this is equivalent to integrating out gravitons in NRGR by using the saddle point approximation to evaluate the path integral, viz., by drawing all tree-level Feynman diagrams. Calculating potentials and power loss in both NRGR and ClEFT boils down to the same set of Feynman rules so using path integrals (which are not directly evaluated anyway) or not does not seem like a significant simplification or modification of NRGR.
Finally, the fourth embodies an attempt to 
apply retarded boundary conditions to the radiation sector of NRGR using only
classical methods and quantitites.
At the orbital scale, perturbations propagate (nearly) instantaneously so there is no difference in using the Feynman or retarded propagators for calculating potentials. Hence, ClEFT is equivalent to NRGR at this scale since both approaches describe the same system in the same classical limit using the same techniques (Feynman diagrams and dimensional regularization) with the same Feynman rules and propagators.  To avoid confusing the reader, we emphasize that in computations to date, ClEFT and NRGR are equivalent and we are reluctant to distinguish ClEFT as an approach that is markedly different from or simpler than NRGR.

However, it is worth pointing out that implementing retarded boundary conditions in a purely classical framework is more involved than stipulating that radiation gravitons represent retarded propagators in Feynman diagrams. Indeed, this Feynman rule is not correct as it leads to an inconsistency.

To see this, we compute the quadrupole radiation. From the third diagram in (\ref{radiation0}) this is found in the TT gauge in ClEFT to be
\begin{eqnarray}
	m_{pl}^{-1} \barh_{\mu\nu}^{TT} = \frac{1}{4 m_{pl}^2} \Lambda _{ij,kl} \int dt^\prime \, D^{ret}_{klmn} (t-t^\prime, \bx) \ddot{Q}_{mn} (t^\prime)  = - \frac{2 G_N}{ | \bx | } \Lambda _{ij,kl} \ddot{Q}_{kl} (t - | \bx | )  ,
	\label{cleftgw10}
\end{eqnarray}
which is the correct expression. However, the quadrupole radiation can be computed directly from the effective action by coupling the radiation graviton to an auxiliary source $J_{\mu\nu} (x^\alpha)$ and taking the variation of $S_{eff}$ so that
\begin{eqnarray}
	\barh_{\mu\nu} (t, \bx) = \frac{ \delta S_{eff} }{ \delta J^{\mu\nu} (t, \bx) } \Bigg|_{J_{\mu\nu} = 0}   .\label{cleftgw0}
\end{eqnarray}
The effective action is calculated by solving the wave equation subject to retarded boundary conditions. Reinserting this solution back into the action, which is what is meant by ``integrating out"  \cite{KolSmolkin:PRD77}, it is easy to show that
\begin{eqnarray}
	S_{eff} = \frac{ 1}{2} \int d^4x \int d^4 x^\prime \, \big[ S^{\mu \nu} (x) + J^{\mu\nu} (x) \big] D^{ret}_{\mu\nu\alpha\beta} (x - x^\prime) \big[ S^{\alpha \beta} (x^\prime) + J^{\alpha \beta} (x^\prime) \big]  \label{cleftgw1}
\end{eqnarray}
where $S_{\mu\nu}$ represents the quadrupole source for the gravitational perturbations and is defined by 
\begin{eqnarray}
	S_{\mu\nu} (t, \bx) \barh^{\mu\nu} (t, \bx) = \frac{Q_{ij} (t) }{2 m_{pl} } R^{0i0j} (t, \bzero) \delta^3 (\bx) .  \label{quadsrc0}
\end{eqnarray}
($S^{\mu\nu}$ should be considered as an operator since the linearized Riemann tensor contains two derivatives acting on the graviton field.)
We assume that the radiation gravitons couple to the auxiliary source bilinearly, $\int d^4x \, J^{\mu\nu} \barh_{\mu\nu}$. Relabeling the integration variables $t \leftrightarrow t'$ allows us to write the retarded propagator in (\ref{cleftgw1}) as the average of the retarded and advanced propagators.
Taking the functional derivative of (\ref{cleftgw1}), as in (\ref{cleftgw0}), gives in the TT gauge
\begin{eqnarray}
	m_{pl}^{-1} \barh_{ij}^{TT} (t, \bx) = \frac{i}{4 m_{pl}^2} \Lambda_{ij,kl} \int dt^\prime \left[ \frac{1}{2} \big( D^{ret} _{klmn} (t-t^\prime, \bzero) + D^{adv} _{klmn} (t-t^\prime, \bzero) \big) \right] \ddot{Q}_{mn} (t^\prime)  ,  \label{cleftgw2}
\end{eqnarray}
which is 
not equal to (\ref{cleftgw10}). 

The Feynman rule for radiation gravitons in ClEFT is therefore seen to be
\begin{eqnarray}
\begin{array}{ccl}
	\mu\nu ~ \parbox{16mm}{\includegraphics[width=16mm]{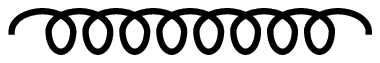}} ~ \alpha \beta & = & \displaystyle\frac{1}{2} \big( D^{ret}_{\mu \nu \alpha \beta} ( x - y) + D^{adv} _{\mu\nu\alpha \beta} (x-y) \big)  
\end{array}
\end{eqnarray}
and is not given by the retarded propagator alone. Indeed, this rule is equivalently given by ${\rm Re}[ i D^F_{\mu\nu\alpha \beta} ]$ and is therefore closely related to the in-out formalism of NRGR. 
As such, it is easy to show that there is no corresponding radiation reaction term in the equations of motion. The problem lies with the fact that the EFT paradigm works at the level of the action and not the equations of motion. Implementing retarded boundary conditions at the level of the Lagrangian requires something different from demanding that all gravitons be retarded solutions to the wave equation. 

Recall that imposing retarded boundary conditions on the radiated perturbations in NRGR effectively amounts to doubling the degrees of freedom such that each set of variables combines in the appropriate way to ensure that the binary's equations of motion and radiated gravitational waves are manifestly real and causal. Motivated by the in-in formalism, we generalize the radiation sector of ClEFT so that radiation reaction and gravitational waves are appropriately derived within a purely classical framework. Indeed, this may be viewed as an alternative derivation of the in-in approach to NRGR.

To make the presentation below clearer we ignore non-linear interactions in the radiation sector. However, their inclusion does not introduce any conceptual obstacles to the framework.

Consider the action for the compact objects and gravitational perturbations that results from doubling the degrees of freedom in a manner analogous to the in-in construction
\begin{eqnarray}
	S [ \{ \bx_{K1} \} , \{ \bx_{K2} \}, \barh_{1,2}^{\mu\nu} ] &=& S [ \{ \bx_{K1} \}, \barh_1 ^{\mu\nu} ] - S [ \{ \bx_{K2} \} , \barh_2 ^{\mu\nu} ]  \label{clefteffaxn0}
\end{eqnarray}
where, in the Lorenz gauge,
\begin{eqnarray}
	S [ \{ \bx_{KA} \} , \barh_A ^{\mu\nu} ] = \frac{1}{2} \int d^4x \left( \partial_\mu \barh_{A\alpha \beta} \partial^\mu \barh_A^{\alpha \beta} -\frac{1}{2} \partial_\mu \barh_A \partial^\mu \barh_A \right)+ \int d^4x \big( J_{A\alpha \beta} + S_{A\alpha \beta} \big) \barh_A^{\alpha \beta}  + O(\barh^3)  ,
\end{eqnarray}
$\barh$ is the trace of the radiation graviton field and $A=1,2$. Here $J_{1,2}^{\mu\nu}$ are auxiliary sources and $S_{1,2}^{\mu\nu}$ denotes the quadrupole source as before. The first term in (\ref{clefteffaxn0}) is the usual action from classical field theory while the second term can be regarded as the action for the system (with $1\to 2$) evolving backward in time. We remark that  (\ref{clefteffaxn0}) is the same action appearing in the CTP path integral of (\ref{genfun2}) with $\bj_{K1} = \bj_{K2} =0$. Also, (\ref{clefteffaxn0}) is not an action typically considered in classical field theory but seems necessary to achieve our purposes here.

Working with the $\barh_\pm ^{\mu\nu}$ variables gives
\begin{eqnarray}
	S [ \{ \bx_{K\pm} \}, \barh_{\pm}^{\mu\nu} ] = \frac{1}{2} \int d^4x \left( \partial_\mu h^a_{\alpha \beta} \partial^\mu h_a^{\alpha \beta} -\frac{1}{2} \partial_\mu \barh^a \partial^\mu \barh_a \right) + \int d^4x  (J^a_{\alpha \beta} + S^a_{\alpha \beta} ) \barh^{\alpha \beta}_a  + O(\barh_\pm^3)   \label{newclefteffS0}
\end{eqnarray}
where the metric $c_{ab}$ is used as before to contract the CTP indices.
Integrating out the gravitational perturbations involves first solving the wave equations
\begin{eqnarray}
	P^{\mu\nu}_{~~~\alpha \beta} \Box \barh_\pm^{\alpha \beta} = J_\pm^{\mu\nu} + S_\pm^{\mu\nu} + O(\barh_\pm^2)  \label{cleftwaveeqn0}
\end{eqnarray}
subject to the appropriate boundary conditions and inserting the solutions back into (\ref{newclefteffS0}). In other words, we extremize the action with respect to variations in the radiation graviton fields and then compute the extremal ``value" of $S$ to obtain the effective action.

As with the in-in formalism, we will set $\bx_{K1} = \bx_{K2} = \bx_K$ and $J_\pm^{\mu\nu}=0$ at the end of all functional variations, which implies that, momentarily dropping the spacetime indices, $\barh_+ \to \barh$ and $\barh_- \to 0$. Notice that the $\barh_+$ equation is sourced by terms that remain non-zero when $\bx_{K1} = \bx_{K2}$. Since $\barh_+$ corresponds to radiated gravitational waves in this limit then these must satisfy retarded boundary conditions so that
\begin{eqnarray}
	\barh_{+\mu\nu} (x) = \int d^4y \, D^{ret}_{\mu\nu\alpha \beta} (x - y) \big[ S_+^{\alpha \beta} (y) + J_+ ^{\alpha \beta} (y) \big]  .   \label{newcleftgw10}
\end{eqnarray}
Solving the $\barh_-$ equation in (\ref{cleftwaveeqn0}) under momentarily unspecified boundary conditions gives
\begin{eqnarray}
	\barh_{-\mu\nu} (x) = \int d^4y \, D_{\mu\nu\alpha \beta} (x-y) \big[ S_-^{\alpha \beta}(y) + J_- ^{\alpha \beta} (y) \big]  \label{newcleftgw11}
\end{eqnarray}
for some propagator $D_{\mu\nu\alpha \beta}$ that will be determined shortly.
Putting these solutions into (\ref{newclefteffS0}) gives the effective action
\begin{eqnarray}
	S_{eff} [ \{ \bx_{K\pm} \} ] &\arreq& \frac{1}{2} \int d^4x \int d^4y \, \big[ S_- ^{\alpha \beta} (x) + J_- ^{\alpha \beta} (x) \big] \big[ D^{ret}_{\alpha \beta \gamma \delta } (x-y) + D_{\alpha \beta \gamma \delta} (y-x) \big] \big[ S_+^{\gamma \delta} (y) + J_+^{\gamma \delta} (y) \big]   + O(\barh_\pm^3) .   \nonumber \\
	&&  \label{newclefteffS1}
\end{eqnarray}
Calculating the variation of the effective action using 
\begin{eqnarray}
	\barh_{\mu\nu} (t, \bx) &=& \barh_{+\mu\nu} (t, \bx) \Big|_{\bx_{K-}=0, \, \bx_{K+} = \bx_K , J_-^{\mu\nu} = J_+^{\mu\nu} = 0 } \\
	 &=& \frac{ \delta S_{eff} }{ \delta J_-^{\mu\nu} (t, \bx) } \Bigg|_{\bx_{K-}=0, \, \bx_{K+} = \bx_K , J_-^{\mu\nu} = J_+^{\mu\nu} = 0 }
\end{eqnarray}
gives the gravitational waves radiated by the binary
\begin{eqnarray}
	\barh_{\mu\nu} (t, \bx) = \frac{1}{2} \int d^4 y \big[ D^{ret}_{\alpha \beta \gamma \delta } (x-y) + D_{\alpha \beta \gamma \delta} (y-x) \big] S^{\gamma \delta} (y; \bx_K) .  \label{newcleftgw0}
\end{eqnarray}
Self-consistency demands (\ref{newcleftgw0}) to equal (\ref{newcleftgw10}) when $\bx_{K1} = \bx_{K2} = \bx_K$ and $J_\pm^{\mu\nu} = 0$. Therefore, $D_{\alpha \beta \gamma \delta} (x-y)$ must be the {\it advanced} propagator. 
With this choice it is straightforward to show that (\ref{newcleftgw0}) is equivalent to (\ref{iningw2}), in the TT gauge, when the $y$-integral is evaluated. 

The fact that $\barh_-$ satisfies {\it advanced} boundary conditions is not a problem because $\bx_{K-}$ and $J_\pm^{\mu\nu}$ are set to zero at the end of the calculations (implying $\barh_-$ vanishes, from (\ref{newcleftgw11})) so that there is no contribution to the equations of motion and gravitational waves from advanced radiation. 
The effective action is therefore
\begin{eqnarray}
	S_{eff} [ \{ \bx_{K\pm} \} ] &\arreq& \int d^4x \int d^4y \, \big[ S_- ^{\alpha \beta} (x) + J_- ^{\alpha \beta} (x) \big] D^{ret}_{\alpha \beta \gamma \delta } (x-y)  \big[ S_+^{\gamma \delta} (y) + J_+^{\gamma \delta} (y) \big]   + O(\barh_\pm^3)   .   \label{newclefteffS2}
\end{eqnarray}
We can write (\ref{newclefteffS2}) in terms of $i$ times the matrix of propagators introduced earlier in the in-in formalism (\ref{propmatrix12})
\begin{eqnarray}
	D^{ab}_{\alpha \beta \gamma \delta } = \left( \begin{array}{cc} 
									0 & D^{adv}_{\alpha \beta \gamma \delta} \\
									D^{ret} _{\alpha \beta \gamma \delta} & 0 
								\end{array} \right)  \label{propmatrix2}
\end{eqnarray}
with $a,b = \pm$ so that (\ref{newclefteffS2}) becomes
\begin{eqnarray}
	S_{eff} [ \{ \bx_{K\pm} \} ] &=& \frac{1}{2} \int d^4x \int d^4y \, \big[ S_a ^{\alpha \beta} (x) + J_a ^{\alpha \beta} (x) \big] D^{ab}_{\alpha \beta \gamma \delta } (x-y) \big[ S_b^{\gamma \delta} (y) + J_b^{\gamma \delta} (y) \big] + O(\barh_\pm^3).  \label{newclefteffS3}
\end{eqnarray}
The CTP indices $a,b$ are raised and lowered with the metric $c_{ab}$ in (\ref{ctpmetric0}). It is straightforward to show that setting the auxiliary sources to zero in (\ref{newclefteffS3}) and varying the effective action with respect to $\bx_{K-}$ gives the radiation reaction force \cite{BurkeThorne:Relativity, Thorne:AstrophysJ158, Burke:JMathPhys12, Maggiore} derived above in (\ref{burkethorne1}). Indeed, (\ref{newclefteffS3}) is simply the in-in effective action of NRGR (\ref{ininrr23}) with factors of $i$ rearranged accordingly.

The formulation of the radiation sector discussed here can be straightforwardly applied to the extreme mass ratio inspiral of a small compact object. 
In this context, there are only radiation gravitons and no potential gravitons to integrate out. Using an appropriate regularization scheme in the presence of a {\it curved} background spacetime we have calculated (but do not present the details here) the well-known self-force on the small compact object  that was first derived in \cite{MinoSasakiTanaka:PRD55, QuinnWald:PRD56}.
Indeed, the calculation is nearly identical (up to factors of $i$ from applying the Feynman rules) to that given in \cite{Galley:EFT1}, which uses the in-in path integral framework.


\bibliographystyle{physrev}
\bibliography{gw_bib}

\setlength{\parskip}{1em}



\end{document}